\documentclass[journal,10pt]{IEEEtran}

\usepackage{cite}
\usepackage{amsmath,amssymb,amsfonts}
\usepackage{algorithmic}
\usepackage{graphicx}
\usepackage{textcomp}
\usepackage{comment}
\usepackage{xcolor}
\usepackage{bbm}
\usepackage{balance}

\def\BibTeX{{\rm B\kern-.05em{\sc i\kern-.025em b}\kern-.08em
    T\kern-.1667em\lower.7ex\hbox{E}\kern-.125emX}}

\DeclareMathOperator*{\argmax}{arg\,max}
\DeclareMathOperator*{\argmin}{arg\,min}

\DeclareMathAlphabet\mathbfcal{OMS}{cmsy}{b}{n}

\newtheorem{thm}{Theorem}

\newtheorem{cor}{Corollary}
\newtheorem{defn}{Definition}
\newtheorem{remark}{Remark}

\begin{document}

\title{Coded Downlink Massive Random Access \\ and a Finite de Finetti Theorem}
\author{
	Ryan Song, \IEEEmembership{Graduate Student Member, IEEE},
	Kareem M. Attiah, \IEEEmembership{Graduate Student Member, IEEE}, \\
        and Wei Yu, \IEEEmembership{Fellow, IEEE}
\thanks{Manuscript submitted to {\it IEEE Transactions on Information Theory} on May 15, 2024, revised on December 25, 2024 and April 8, 2025. The authors are with The Edward S.\ Rogers Sr.\ Department of Electrical and Computer Engineering, University of Toronto, Canada. (E-mails:  r.song@mail.utoronto.ca, kattiah@ece.utoronto.ca, weiyu@ece.utoronto.ca.) The materials in this work have been presented in part at 
IEEE International Symposium on Information Theory, June 2022, Helsinki, Finland \cite{categorization} 
and in part at IEEE International Symposium on Information Theory, July 2023, Taipei, Taiwan \cite{coded-downlink}.
This work was supported by the Natural Sciences and Engineering Research Council (NSERC) of Canada via a Discovery Grant and a Post-Graduate Scholarship. 
}
}
\maketitle
\thispagestyle{empty}

\begin{abstract}
This paper considers a massive connectivity setting in which a base-station
(BS) aims to communicate sources $(X_1,\cdots,X_k)$ to a randomly activated
subset of $k$ users, among a large pool of $n$ users, via a common message
in the downlink. Although the identities of the $k$ active users are assumed to be
known at the BS, each active user only knows whether itself is active and does
not know the identities of the other active users. A naive coding strategy is
to transmit the sources alongside the identities of the users for which
the source information is intended. This requires $H(X_1,\cdots,X_k) + k\log(n)$ bits, 
because the cost of specifying the identity of one out of $n$ users is $\log(n)$ bits. 
For large $n$, this overhead can be significant. This paper shows that it
is possible to develop coding techniques that eliminate the dependency of the overhead 
on $n$, if the source distribution follows certain symmetry. Specifically,
if the source distribution is independently and identically
distributed (i.i.d.) then the overhead can be reduced to at most $O(\log(k))$ bits, 
and in case of uniform i.i.d.\ sources, the overhead can be further reduced to $O(1)$ bits. 
For sources that follow a more general exchangeable distribution, 
the overhead is at most $O(k)$ bits, and in case of finite-alphabet exchangeable sources, 
the overhead can be further reduced to $O(\log(k))$ bits. The downlink massive random 
access problem is closely connected to the study of finite
exchangeable sequences. The proposed coding strategy allows bounds on the
Kullback-Leibler (KL) divergence between finite exchangeable distributions and i.i.d.\
mixture distributions to be developed and gives a new KL divergence version
of the finite de Finetti theorem, which is scaling optimal. 
\end{abstract}

\begin{IEEEkeywords}
Massive random access, finite de Finetti theorem, exchangeable distribution
\end{IEEEkeywords}

\section{Introduction}
\label{section:introduction}

This paper considers the problem of information transmission from a central
base-station (BS) to a random subset of $k$ users in a downlink
massive random access scenario with a large pool of $n$ users. The users are
assumed to be sporadically activated, so that at any given time, only $k \ll n$
users are actively listening to the BS. The BS knows the identities of the
active users and wishes to communicate sources $(X_1, \cdots, X_k)$ to these
active users via a downlink common message; but each user only knows whether it
is active itself and does not know which other users are active. We ask the
question: What is the minimum length of the common message that allows
each active user $i$ to learn the source $X_i$ intended for it?

\subsection{Motivation and Problem Setting}

The aforementioned source coding problem, which we refer to as the \emph{coded downlink
massive random access} problem, arises naturally in the context of machine-type
communications or Internet-of-Things (IoT), particularly in scenarios where the
users are randomly activated and the BS needs to send information (such as the
acknowledgment or control messages) to the active users in the downlink
\cite{amp-activity-detection, cov-activity-detection}.  
A typical communications protocol in this context consists of two phases. 
In the first phase, the active users transmit pilots
to the BS, and the BS performs activity detection to learn the identities of
the active users.  In the second phase of the protocol, the BS transmits the
information sources to each of the active users via a downlink common message.
This paper makes the assumption that the user activities are detected correctly
in the first phase and focuses on the fundamental limit of communications
in the second phase.

The sources intended for the active users are not necessarily
independent (e.g., there may be a cap on the maximum number of users that can be
positively acknowledged, or the message may convey the assignment of non-colliding 
scheduling slots to the active users), but under reasonable assumptions, the distribution of the 
sources typically possesses certain symmetry. In this paper, we assume that the user activity
patterns are random and symmetric across the pool of all potential users, i.e.,
no subset of users is preferred over any other subset. Moreover, we assume that 
the sources intended for the active users are independent of the user activity pattern.
Furthermore, fixing a user activity pattern, we assume that all permutations of 
any source sequence are equally probable. 
This symmetry under permutations is known as \emph{exchangeability}.

Formally, sources $(X_1, \cdots, X_k)$ are \emph{exchangeable} if their distribution
$p(x_1, \dots, x_k)$ is invariant under permutation, i.e. for any bijection $\sigma$ 
from $\{1,2, \cdots, k\}$ to itself, we have
\begin{equation}
    p(x_1, \cdots, x_k) = p(x_{\sigma(1)}, \cdots, x_{\sigma(k)}).
\end{equation}
Classical examples of exchangeable distributions include independently and
identically distributed (i.i.d.) random variables, i.i.d.\ mixture random
variables, and urn distributions (i.e., sampling with or without
replacement from a population). 

Throughout this paper, the total number of potential users $n$ is assumed to be
known and fixed. We begin the exposition by assuming that the number of active
users $k$ is also fixed, then relax this assumption for i.i.d.\ sources and 
treat the case of random $k$ in a later section.

The downlink massive random access problem would have been trivial if every
active user knew the identities of the entire subset of active users. In this
case, the $n$ users can each be pre-assigned an index. In the downlink phase, 
the BS can then list the sources for the $k$ active users according to the
order in which they appear in the index. Such a common message requires only
$H(X_1,\cdots,X_k)$ bits, but relies on each user knowing the identities of all
active users. 

The problem becomes much more interesting under the practical scenario in which
every active user only knows whether it is active itself and does not know who
else is active. In this case, it would appear that the BS needs to send not
only $H(X_1,\cdots,X_k)$ bits for the sources, but also a header of $\log(n)$
bits per active user to describe which user each source is intended for, thus resulting
in an overhead of $k\log(n)$ bits. Such an overhead can be
significant when $n$ is large (e.g., $n=10^6$ with $k=10^3$ would require
$k\log(n) \approx 20$ kilobits), especially if the payload for each user is small
in comparison. The main insight of this paper is that if the sources are
exchangeable, then it is possible to reduce the overhead significantly as 
compared to the aforementioned naive scheme. In fact, it is possible to 
develop a coded downlink massive random access strategy with a
common message length that does not depend on $n$.

The intuition that makes this possible is that the naive scheme broadcasts 
too much redundant information. Specifically, each active user needs to recover
only its own designated source, and does not care which other users are
active, or the sources intended for the other users. The naive scheme
broadcasts the source intended for each of the active users to everyone. 
In contrast, the coding strategy developed in this paper reduces the common
message length by taking advantage of the fact that only the $k$ active users
are listening and that each of the $k$ active users is only interested in the
source pertaining to itself.  This allows a single codeword to cover many
different instances of the activity patterns and the associated sources.
Equivalently, it allows the common message length needed to convey the $k$ sources
to a random subset of $k$ active users to be significantly reduced as compared 
to the naive scheme.

\subsection{Proposed Coding Scheme}

This paper proposes a coding strategy using a codebook comprised of length-$n$
codewords of different possible realizations of the sources. Each of the $n$
users has a corresponding unique location in the codewords, so that when the
identities of the $k$ activity users and their associated sources are revealed,
the BS can communicate the sources to the active users by searching 
over the codebook and broadcasting the index of the first codeword that \emph{matches} 
the sources intended for the $k$ active users. 

\begin{figure}
    \centering
    \includegraphics[width=0.3\textwidth]{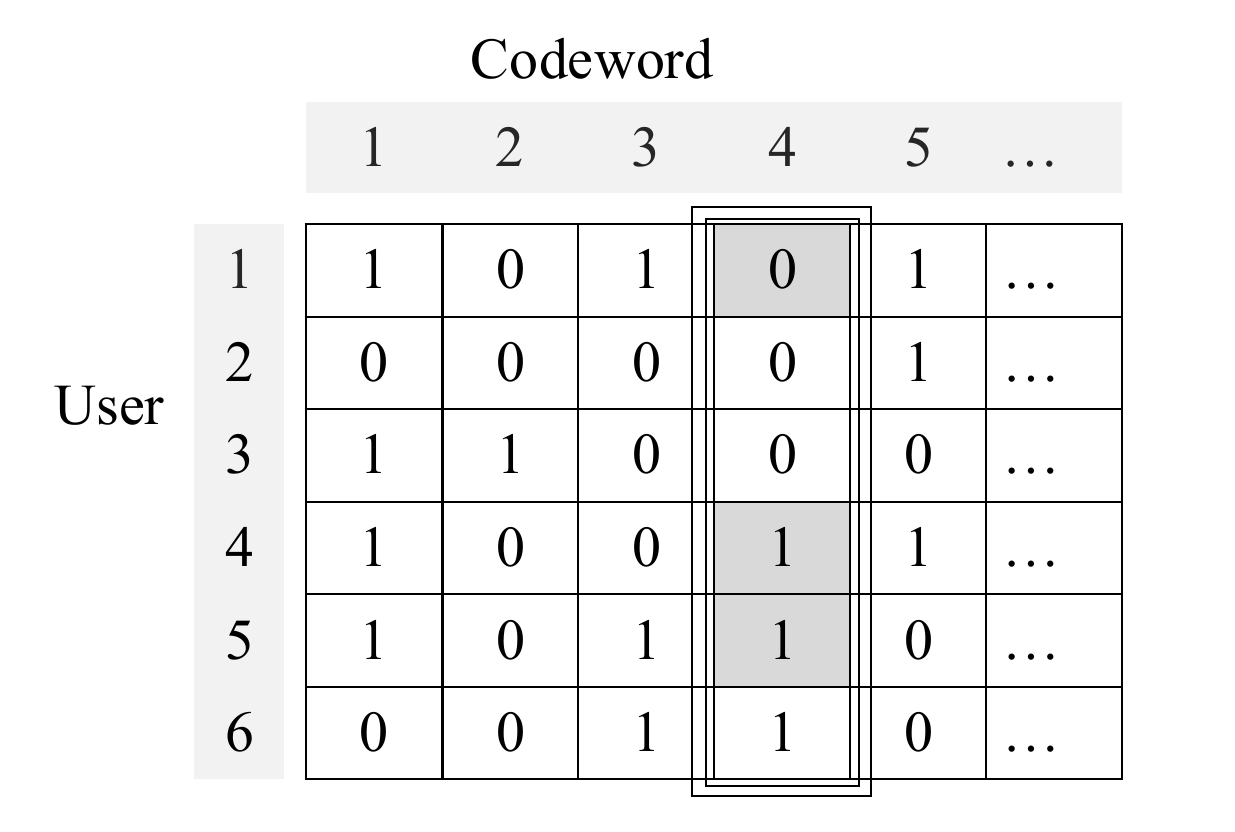}
    \caption{\label{fig:example-binary}{\color{black}Codebook for transmitting i.i.d. Ber$(\frac{1}{2})$ sources to $k=3$ active users among a total of $n=6$ potential users.}} 
\end{figure}


As an example, suppose that we wish to communicate i.i.d.
$\mathrm{Ber}(\frac{1}{2})$ sources to $k=3$ active users among a total of $n=6$
potential users. The source entropy is $k$ bits. Instead of using $k + k\log(n)$ bits
to send one bit to each of the $k$ active users,
we use a codebook comprised of length-$n$
binary vectors, as illustrated for this example in Fig.~\ref{fig:example-binary}, 
where each of the $k$ active users has a unique location in the codewords. 
Suppose that the BS wishes to communicate binary sources $(0,1,1)$ to active users
$(1,4,5)$, respectively. The BS would search through the codebook as shown 
in Fig.~\ref{fig:example-binary}, select the codeword $4$, (because it is the codeword 
with the smallest index that has the values $(0,1,1)$ at the active user 
locations $(1,4,5)$), and broadcast this index $4$ to allow all the active users 
to recover the sources intended for them. Note that in this coding
scheme, a single codeword can be used to communicate many different
source-activity pairs. This is because only the active users are listening and the codeword values at the locations of the non-active users are irrelevant. For example, codeword $4$ would also suffice for
communicating sources $(0,0,1)$ to active users $(2,3,4)$.

If the user activity patterns are random and the sources are i.i.d.\
$\mathrm{Ber}(\frac{1}{2})$, we can construct a random codebook with 
i.i.d.\ $\mathrm{Ber}(\frac{1}{2})$ entries. In this case, the index of 
the first codeword that matches the sources at the active users' locations 
follows a geometric distribution with parameter $p = \left(\frac{1}{2}\right)^k$. 
Let random variable $T$ denote this index, we have 
\begin{equation}
    \mathrm{Pr}(T=t) = (1-p)^{t-1} p.
\end{equation}
The entropy of a geometric random variable is bounded from above as 
\begin{eqnarray}
    H(T) &=& \frac{-p\log(p) - (1-p)\log(1-p)}{p} \\
    &\leq& \log \left( \frac{1}{p} \right) + \log(e) \\
    &=& k + \log(e).
\end{eqnarray}
Thus by using an optimal prefix-free code for compressing $T$, the expected 
common message length can be shown to be within $\log(e)+1$ bits of the joint entropy
of the sources, which is $k$ bits. Most importantly, this same coding strategy 
works regardless of $n$. In fact, the entropy of the index does not depend on $n$; 
this contrasts the naive scheme, which would have required an overhead of $k \log(n)$ bits. 

The main result of this paper is that the coding strategy illustrated in Fig.~\ref{fig:example-binary} can be generalized to arbitrary exchangeable sources. When the sources are exchangeable, it is possible to design a codebook so that the entropy of the matching index is $H(X_1,\cdots,X_k)$ plus an overhead independent of $n$.

\subsection{Connection with Finite de Finetti Theorem}

This paper further points out that the problem of constructing a good codebook
for exchangeable sources is closely related to the de Finetti theorem and
the study of finite exchangeable sequences. This connection arises due to the
fact that the codebook construction needs to be symmetric across all subsets of
users. Therefore, it is natural to use an i.i.d. mixture distribution to construct the codebook, i.e., each codeword is generated according to the distribution 

{\color{black}
\begin{equation}
    q(x_1, \cdots x_n) = \int_{\theta} r(\theta)\left(\prod_{i=1}^{n} q(x_i|\theta)\right)d\theta.
\end{equation}
}

The distribution of the sources however can be any
exchangeable distribution. Therefore, the problem of constructing a good
codebook for exchangeable sources reduces to finding an i.i.d.\ mixture
distribution which is close to the exchangeable distribution in terms of Kullback-Leibler (KL) divergence. In this paper, we propose a codebook construction method for
exchangeable sources and utilize it to prove a KL divergence version of
the finite de Finetti theorem which matches the optimal scaling 
of \cite{dia-fre}.

\subsection{Related Work}

The coding strategy presented in this paper takes inspiration from the
collision-free scheduling codes of \cite{scheduling}. In the problem of
collision-free scheduling, the BS aims to assign each of the $k$ active users
into one of $k$ transmission slots, such that no two active users are assigned
to the same slot. Since the BS is not assigning users to specific slots, this
problem does not directly correspond to the coded downlink transmission of
information sources. Later in this paper, we 
discuss an alternative scheduling problem in which BS wishes to assign each 
active user to a specific transmission slot. This alternative scheduling 
problem would be an instance of the coded downlink transmission. Although the original
problem formulation in \cite{scheduling} is different from the setting of this
paper, the code constructions bear similarities. Both leverage the fact that a single
codeword can be used to schedule many different sets of active users.


It is worth noting that the scheduling code from \cite{scheduling} in
conjunction with conventional source coding already allow a coding strategy for
exchangeable sources at common message lengths independent of $n$. The idea is 
that we can first schedule the $k$ active users into $k$ distinct slots using at most
$(k+1) \log(e)$ bits using the strategy in \cite{scheduling}, then list the
sources in the order of the slot assignments, so that each active user receives the correct information. Since the sources are exchangeable, the distribution is not affected by the reordering. Therefore a common message length of $H(X_1, \cdots, X_k) + (k+1)\log(e)$ bits is achievable.
This strategy achieves the same $O(k)$ overhead as the coding scheme proposed in
this paper for general exchangeable sources, but in many cases, the $O(k)$ overhead can 
be improved. For example, if the sources are i.i.d.\ or i.i.d.\ mixture, the coding 
strategy proposed in this paper achieves an overhead of $O(\log(k))$ bits.
For sources with uniform distribution over its alphabet, the overhead can be further reduced to $O(1)$ bits.

The problem setting of this paper, when specialized to transmitting i.i.d.\ bits, is reminiscent of the concept of \emph{identification via channels} 
\cite{ahlswede-dueck} and specifically the idea of \emph{identification plus transmission} 
(IT) code in \cite{han-verdu}, which aims to send a message to \emph{one} intended user
among a large pool of potential users via $n$ uses of a channel of capacity $C$, so that 
the intended user receives the message with error probability at most $\lambda_1$, while 
the average successful decoding probabilities at all other non-intended users are upper bounded by $\lambda_2$.
The key result of \cite{han-verdu} is that for any positive $(\lambda_1,\lambda_2)$, there 
exists a fixed-length code for the setting in which the number of messages can be up to $2^{nC}$ 
and the number of potential users can be up to $2^{2^{nC}}$.
In other words, the ``addressing'' information for identifying the intended user in a pool of up to
doubly exponential number of potential users can be embedded into the message for the intended user 
at essentially no extra cost in rate. 

The problem setting of this paper differs from that of \cite{han-verdu} in that: (i) we consider the case of $k$ intended recipients 
rather than a single user; (ii) we enforce zero error for the intended users, rather than allowing
some positive error probability $\lambda_1$; (iii) we assume that all the other users are not listening,
thus we do not consider the error probability $\lambda_2$; but the number of potential users can
be unbounded, rather than being doubly exponential; (iv) we consider variable-length coding instead of fixed-length codes. 
We show that the ``addressing'' overhead in the setting of this paper for transmitting i.i.d.\ 
Ber$(\frac{1}{2})$ bits to $k$ users is at most a constant, regardless of the size of the pool
of potential users. 

The coded downlink information transmission problem for massive random access 
has been considered in the earlier conference version of this work \cite{categorization, 
coded-downlink}, where a specific example of the categorization problem and 
its generalization are introduced.  In this paper, we expand upon
\cite{categorization, coded-downlink} to treat more general sources. 
Furthermore, we draw a connection between the massive random access problem and the theory of
finite exchangeable sources. 


Finite de Finetti theorems have a long history \cite{stam, dia-fre, matus}.
Let $(X_1, \cdots, X_k)$ be an exchangeable sequence with distribution $p(\mathbf{x})$ which takes values in $\mathcal{X}^k$ and let $\mathcal{Q}$ be the set of all i.i.d.\ mixture distributions on $\mathcal{X}^k$. We say that $(X_1, \cdots, X_k)$ is \emph{$d$-extendable} if it has the same distribution as the first $k$ elements of a longer exchangeable sequence $(X_1, \cdots, X_k, \cdots, X_d)$. {\color{black}From \cite[Theorems $3$ and $13$]{dia-fre}}, for all $d$-extendable exchangeable sequences $(X_1, \cdots, X_k)$ taking values in $\mathcal{X}$, we have
\begin{equation}
\label{eq:intro-dia-fre}
    \min_{q \in \mathcal{Q}} \mathrm{TV}(p, q) \leq \min \left\{ \frac{k(k-1)}{2d}, \frac{k|\mathcal{X}|}{d} \right\},
\end{equation}
where $\mathrm{TV}(p, q) = \frac{1}{2} \sum_{\mathbf{x} \in \mathcal{X}^k} | p(\mathbf{x}) - q(\mathbf{x}) |$ is the total variation (TV) distance\footnote{This definition of TV distance differs from the definition used in \cite{dia-fre} by a factor of $1/2$.}. Therefore for any fixed $k$, the TV distance tends to $0$ as $d \to \infty$. In fact, the TV distance can go to $0$ even if $k \to \infty$, as long as $\frac{k}{d} \to 0$ for finite $|\mathcal{X}|$, or $\frac{k^2}{d} \to 0$. It is shown in \cite{dia-fre} that this scaling is optimal. 

{\color{black}
Recently, there has been an interest in finding similar bounds as \eqref{eq:intro-dia-fre} but in terms of KL divergence instead of TV distance. In particular, \cite{exch-it-1, exch-it-2, exch-it-3, exch-it-4} utilize information-theoretic arguments to prove various upper bounds on the KL divergence. It is shown in \cite{exch-it-3} that
\begin{equation}
\label{eq:ub-with-entropy}
    \min_{q \in \mathcal{Q}} D(p\|q) \leq \frac{k(k-1)}{2(d-k+1)} H(X_1),
\end{equation}
where $D(p\|q)$
is the KL divergence. This bound holds for sources with arbitrary alphabet sizes, but the bound depends on the distribution of the sources. In the finite alphabet setting, \cite{exch-it-4} utilizes the log-sum inequality along with results of \cite{stam} to show that
\begin{equation}
\label{eq:stam-arg}
    \min_{q \in \mathcal{Q}} D(p\|q) \leq (|\mathcal{X}|-1)\frac{k(k-1)}{2(d-1)(d-k+1)}.
\end{equation}
Although the scalings of \eqref{eq:ub-with-entropy} and \eqref{eq:stam-arg} match that of \cite{dia-fre}, both can be improved upon. This paper improves \eqref{eq:ub-with-entropy} by removing the dependency on the distribution of the sources. 
Further, in the case of $d=k$, the bound \eqref{eq:ub-with-entropy} implies a dependency of $O(k^2)$, while the result of this paper improves the bound to $O(k)$.  
Moreover, when the alphabet size is finite, 
in the case of $d=k$, the bound \eqref{eq:stam-arg} implies a dependency of $O(|\mathcal{X}|k)$. In this paper, we tighten this upper bound to $O(|\mathcal{X}|\log(k))$ using a method-of-types argument.
}
\subsection{Main Contributions}

The main technical results of the paper are as follows:

\subsubsection{Coding for Downlink Massive Random Access}

For transmitting exchangeable sources $\mathbf{X} = (X_1, \cdots, X_k) \sim p(\mathbf x)$ to a random subset of $k$ active users in a downlink massive random access scenario with $n$ total users, we show that a common message length of
\begin{equation}
    \min_{q \in \mathcal{Q}} H(\mathbf{X}) + D(p\|q) + \log(H(\mathbf{X}) + D(p\|q) + 1) + 2
\end{equation}
bits is achievable, where $\mathcal{Q}$ is the family of i.i.d.\ mixture distributions on $\mathcal{X}^k$.

In the case that the sources are i.i.d.\ or i.i.d.\ mixture, we can set $q(\mathbf{x}) = p(\mathbf{x})$ to eliminate the $D(p\| q)$ term. Specifically, for i.i.d.\ sources, an achievable common message length is $kH(X_1)+\log(k H(X_1)+1)+2$, thus the overhead is at most $O(\log(k))$ bits. The overhead can be reduced to $O(1)$ if the sources are uniform over the alphabet in its support. 

For general exchangeable sources that are not i.i.d.\ or i.i.d.\ mixture, we show that
\begin{equation}
\label{eq:intro-exch}
    \min_{q \in \mathcal{Q}} D(p\|q) \leq \min\{k\log(e), |\mathcal{X}|\log(k+1) \},
\end{equation}
thus achieving an overhead of at most $O(k)$ bits in general and $O(\log(k))$ bits if the source has a finite alphabet. 

In all cases, the achievable common message length is close to $H(\mathbf{X})$ bits with an overhead that does not depend on $n$. 


\subsubsection{Extendable Sources and Finite de Finetti Theorem}

The upper bound \eqref{eq:intro-exch} can be improved if the sources $\mathbf{X} \sim p(\mathbf{x})$ are $d$-extendable. 
In this case, we show that for $d > k$
\begin{equation}
\label{eq:intro-ext}
    \min_{q \in \mathcal{Q}} D(p\|q) \leq \min \left\{ \log\left(\frac{d^k}{d^{\underline{k}}} \right), (|\mathcal{X}|-1) \log \left(\frac{d-1}{d-k} \right) \right\},
\end{equation}
where $d^{\underline{k}} = d(d-1)\cdots(d-k+1)$. 
This is a statement of the finite de Finetti theorem in terms of KL divergence. 

This bound is independent of the distribution of the sources.
Further, by using the fact $ \log \left( \frac{d^k}{d^{\underline{k}}} \right) \leq - \log \left(1 - \frac{k(k-1)}{2d} \right)$
for $d > \frac{1}{2}k(k-1)$, it can be shown that 
the KL divergence between the
distribution $p(\mathbf{x})$ and the nearest i.i.d.\ mixture distribution goes to $0$ 
as long as $\frac{k^2}{d} \to 0$ for general alphabet case and $\frac{k}{d} \to
0$ in the finite alphabet case.  
These scaling results match the optimal scalings of \eqref{eq:intro-dia-fre}. 


\subsection{Paper Organization and Notation}

The remainder of the paper is organized as follows. Section
\ref{section:lossless} states the problem of coded downlink massive random access 
and develops the achievability results. 
The improved finite de Finetti theorem is stated and proved in 
Section \ref{section:finetti}. 
Conclusions are drawn in Section \ref{sec:conclusion}.


Throughout this paper, we use lowercase letters to denote scalars, lowercase boldface letters to denote vectors, capital letters to denote random variables, boldface capital letters to denote random vectors, and calligraphic letters, e.g. $\mathcal{S}$, to denote sets, $|\mathcal{S}|$ to denote their cardinality. 
We let $\log(\cdot)$ denote the base 2 logarithm and $\ln(\cdot)$ denote the natural logarithm. All information measures are expressed in bits, including entropy $H(\cdot)$ and the KL divergence $D(\cdot\|\cdot)$. We use $\mathbb{N}$ to denote the set of natural numbers and $\mathbbm{1}(\cdot)$ as the indicator function. We use Ber$(p)$ to denote Bernoulli random variable with parameter $p$. We use the shorthand notations $[n] = \{ 1,\dots,n \}$ and $a^{\underline{b}} = a(a-1)\cdots(a-b+1)$.




\section{Coding for Downlink Massive Random Access}
\label{section:lossless}

\subsection{Problem Formulation} 
Consider a massive random access setting in which a random subset of $k$ users
become active among a total number of $n$ users. The value of $n$ is fixed and
known. For now, $k$ is also assumed to be fixed and known.
The identities of the active users are known to the BS.  However, each user
only knows whether itself is active, but does not know which other users are active.
We are interested in the setting where the BS wishes to
communicate a source to each of the $k$ active users simultaneously using a
common message in the downlink. This common message is assumed to be received by all users without error.

In this paper, we consider the class of sources that are
exchangeable. Let $\mathbf{X} = (X_1, \cdots, X_k)$ be a sequence of
exchangeable random variables taking values from a finite or countable alphabet $\mathcal{X}$. 
Let $p(\mathbf{x}) = p(x_1,\cdots,x_k)$ be their joint distribution. 
Each user $i$ is interested in learning $X_i$ based on the common message (but
not interested in any of other $X_i$'s). 

The following are some common applications in which exchangeable distributions may arise:
\begin{itemize}
\item[1)] The BS wishes to communicate i.i.d.\ $X_i \sim p(x)$.
\item[2)] The $k$ active users need to be assigned into $b$ slots, where $b \ge k$, in a non-colliding fashion, i.e., no two users can be assigned to the same slot. This is known as scheduling. Fixing a particular schedule, the BS wishes to communicate to each user which slot it is assigned to. 
\item[3)] The $k$ active users need to be categorized into $c$ categories where each category has a fixed number of $k_\ell$ users and $\sum_{\ell = 1}^{c} k_{\ell} = k$. 
The BS wishes to communicate the category labels $X_i \in [c]$ to each user.  
For the case of two categories ($c=2$), this is known as the acknowledgment problem.
\item[4)] A fixed number of resources $r$ need to be distributed among the $k$ active users, i.e., $\sum_{i=1}^{k}X_i = r$, where $X_i \ge 0$ is an integer. The BS wishes to communicate to each user how many units of resources it has been assigned. This is known as the resource allocation problem.
\end{itemize}

Let the random variable $\mathbf{A} \in \mathcal{A}^{(n,k)}$ denote the identities of
the $k$ active users, where 
\begin{equation}
 \mathcal{A}^{(n,k)} = \{ \mathbf{a} \in [n]^k \ | \ a_i \neq a_j, \forall i \neq j\}.
\end{equation}
Here, $a_i \in [n]$ is the index of the $i$th active user. 
While $\mathbf{X}$ describes the source contents, the activity pattern $\mathbf{A}$ indicates which users should receive which source, i.e., 
we want each user $a_i$ to receive $x_i$ $\forall i \in [k]$. 
Together, they form a source-activity pair $(\mathbf{X}, \mathbf{A})$. Throughout this paper, we assume that $\mathbf{X}$ and $\mathbf{A}$ are independent.
Notationally, we use $(\mathbf{x}, \mathbf{a})$ to represent a realization of $(\mathbf{X}, \mathbf{A})$.

The problem of communicating the sources $\mathbf X$ to the active users in $\mathbf A$ can now be thought of as a one-shot source coding problem consisting of a single encoder and multiple decoders. {\color{black} The BS uses an encoder $f$ to map a source-activity pair $(\mathbf{x}, \mathbf{a})$ to a binary string, i.e., 
\begin{equation}
\label{eq:problem-formulation-encoder}
  f: \mathcal{X}^k \times \mathcal{A}^{(n,k)} \rightarrow \{0,1\}^*.
\end{equation}
It is assumed that the binary string is sent to all the active users in an error-free fashion. 
Each active user then uses its decoder \color{black} $d_{a_i}: \{0,1\}^* \to \mathcal{X}$ to recover its intended source based on the binary string. We require lossless recovery of the sources: 
\begin{equation}
\label{eq:encoder-decoder-correctness}
 d_{a_i}(f(\mathbf{x}, \mathbf{a})) = x_{i},  \ \ \forall (\mathbf{x}, \mathbf{a}) \in \mathcal{X}^k \times \mathcal{A}^{(n,k)}, \ \forall i \in [k],
\end{equation}
i.e. each active user must receive its intended source without error. Note that users are not required to learn anything pertaining to the other active users, whether that be their identities or the intended sources for the other users.

To ensure unique decodability, we require the set of all possible output binary strings 
of the encoder to be a \emph{prefix-free code}. The optimal encoder and decoders 
$(f^*(\mathbf{x},\mathbf{a}), d_{a_1}^*,\cdots d_{a_k}^*)$,
$(\mathbf{x}, \mathbf{a}) \in \mathcal{X}^k \times \mathcal{A}^{(n,k)}$
are defined to be the encoder and decoders that minimize $\mathbb{E}[\mathtt{len}(f(\mathbf{X}, \mathbf{A}))]$ with $\mathtt{len}(\cdot)$ denoting the length of a string, while satisfying the condition \eqref{eq:encoder-decoder-correctness}. 
The optimal common message length is 
\begin{equation}
	R^* \triangleq \mathbb{E}\left[\mathtt{len}\left(f^*(\mathbf{X}, \mathbf{A})\right)\right],
	\label{R_star}
\end{equation}
where $f^*(\mathbf{X}, \mathbf{A})$ is the optimal encoder.

}

{\color{black}
\begin{remark}
This paper focuses on the setting in which the downlink channel is noiseless and has
the same capacity to all users. This turns the design of optimal common message
encoder and decoders into a source coding problem. If instead, the downlink
channel is noisy, then either appropriate channel coding can be used to turn
the downlink channel into an error-free channel, or if the problem is analyzed from the point of view of transmitting correlated sources over a noisy broadcast channel (e.g., \cite{js-coding-broadcast}), it would lead to a highly complex problem setting, 
which we will not treat in this paper.
\end{remark}

}

\begin{remark}
The definitions of the encoder \eqref{eq:problem-formulation-encoder} and the expected
codeword length \eqref{R_star} assume that the common message can have variable length.
Although in many communications context, fixed-length codes may be more desirable at a protocol level, the optimal fixed-length codes are considerably more difficult to construct 
for this downlink massive random access problem. The existence of fixed-length codes has 
been shown for specific problem instances such as collision-free scheduling \cite{scheduling}, categorization \cite{categorization}, and user acknowledgment \cite{arq-feedback}, 
but the proofs often involve the use of probabilistic arguments and are non-constructive \cite{erdos}. 
Further, if the source alphabet is infinite, fixed-length codes cannot exist. For these
reasons, the rest of this paper focuses on code construction for variable-length common 
messages.
\end{remark}

\subsection{Codebook Construction}

{\color{black}

This paper proposes an encoding and decoding strategy involving two stages.
The BS first uses a function $g$ to map a source-activity pair to a positive index, i.e., 
\begin{equation}
  g: \mathcal{X}^k \times \mathcal{A}^{(n,k)} \rightarrow \mathbb{N},
\end{equation}
then uses an optimal prefix-free code to compress the output of $g(\mathbf{x}, \mathbf{a})$ into a variable-length binary string $f(\mathbf{x},\mathbf{a})$, which is subsequently broadcast to all active users. On the decoding side, each active user $a_i$ first decodes $g(\mathbf{x}, \mathbf{a})$ based on the received binary string, then recovers their respective sources. 
}

Specifically, the proposed encoding and decoding scheme utilizes a shared codebook between the BS and all the users
\begin{equation}
\label{eq:codebook-m}
  \mathbf{m} = (\mathbf{c}^{(1)}, \mathbf{c}^{(2)}, \cdots),
\end{equation}
which consists of a {\color{black}potentially infinite} sequence of length-$n$ vectors $\mathbf{c}^{(t)} \in
\mathcal{X}^n$. 
We assign a unique entry location in the codewords to each of the $n$ users.
For a particular source-activity tuple $(\mathbf{x},\mathbf{a})$, the BS encodes 
$(\mathbf{x},\mathbf{a})$ by finding a codeword $\mathbf{c}^{(t)}$ such that every 
active user has the correct source content in their designated entry in $\mathbf{c}^{(t)}$, i.e.,
\begin{equation}
c^{(t)}_{a_i} = x_i, \qquad \forall i \in [k]. 
\end{equation}
%
We construct the codebook $\mathbf{m}$ so that such a matching codeword always exists. 
If there are multiple matching codewords, we define the encoder output
to be the index of the \emph{first} matching codeword in the codebook.

Denoting $g_{\mathbf{m}}(\mathbf{x},\mathbf{a})$ as the encoder output for the source-activity pair $(\mathbf{x},\mathbf{a})$ using the codebook $\mathbf{m}$, we have
\begin{eqnarray}
\label{eq:encoder}
g_\mathbf{m}(\mathbf{x}, \mathbf{a}) = & \min &  t \\
& \mathrm{s.t.} & {c^{(t)}_{a_i} = x_i , \quad \forall i \in [k]}. \nonumber
\end{eqnarray}
This output is further compressed using a prefix-free code. 
Since any random variable can be compressed using an optimal prefix-free code into a binary string with an expected length within one bit of its entropy, the common message length of the proposed encoding scheme is bounded by 
\begin{equation}
R < H(g_\mathbf{m}(\mathbf{X}, \mathbf{A})) + 1.
\end{equation}
At the decoder, each user decompresses the prefix-free code, thereby recovering $g_\mathbf{m}(\mathbf{x}, \mathbf{a})$ or equivalently the minimum $t$.
The decoding process for each user $u \in [n]$ is simply
\begin{equation}
\label{eq:decoder}
    d_u(t) = c^{(t)}_{u}.
\end{equation}
One can easily verify that this encoding and decoding scheme 
satisfies the condition \eqref{eq:encoder-decoder-correctness}. 

It remains to discuss how to generate the codewords in $\mathbf{m}$. 
We use a random codebook construction in
which the codewords are generated independently according to a mixture of
i.i.d.\ distributions. Specifically, fix a distribution $r(\theta)$ and a
conditional distribution $q(\mathbf x|\theta)$. 
Each length-$n$ codeword $\mathbf{c}^{(t)} = [x_1,\cdots,x_n]^\top$ is generated according to 
\begin{equation}
    q(x_1, \cdots x_n) = \int_{\theta} r(\theta)\left(\prod_{i=1}^{n} q(x_i|\theta)\right)d\theta.
\label{eq:codebook_generation}
\end{equation}
Operationally, each codeword can be thought of as being generated in a two-step process of first choosing a $\theta$ according to $r(\theta)$, then generating the entries of the codeword in an i.i.d.\ fashion according to $q(\mathbf x|\theta)$. This makes each entry of the codeword conditionally i.i.d. Each subsequent codeword is generated independently in the same way and the entire codebook is shared between the BS and all users. 


With this random codebook construction, the distribution of any $k$ distinct entries of a single codeword is
\begin{equation}
\label{eq:q-distrib}
    q(x_1, \cdots x_k) = \int_{\theta} r(\theta)\left(\prod_{i=1}^{k} q(x_i|\theta)\right)d\theta.
\end{equation}
This means that the distributions of any $k$ distinct entries of the codewords
are identical, which is a desired property for encoding exchangeable sources.
This distribution can be specifically designed according to any i.i.d.\
mixture.  

This paper focuses attention to the encoding and decoding functions and random
codebooks generated in this way. The main result of this paper is that for any
exchangeable distribution $p(x_1,\cdots,x_k)$, we can choose an
appropriate i.i.d.\ mixture distribution $q(\mathbf x|\theta) r(\theta)$, such that
among an ensemble of codebooks randomly generated according 
to (\ref{eq:codebook_generation}), there exists a codebook $\mathbf{m}^*$ that 
has a corresponding $H(g_{\mathbf{m}^*}(\mathbf{X}, \mathbf{A}))$ upper 
bounded by $H(X_1,\cdots,X_k)$ plus a small overhead. 


{\color{black}
\begin{remark}
As an alternative to sharing a codebook of potentially infinite size, it is also possible for the BS and the users to utilize common randomness to generate random codewords as they are needed. In a sense, the codewords can be thought of as the output of a random hash function from the user indices $[n]$ to the source alphabet $\mathcal{X}$ (according to the distribution \eqref{eq:codebook_generation}). In this paper, we assume either the availability of common randomness or the ability to share an infinite codebook between the BS and the users.
\end{remark}

\begin{remark}
Throughout this paper, we assume that there are no errors in the activity
detection phase, i.e., the BS knows the identities of the active users
perfectly.  If instead, some inactive users are incorrectly detected as active,
the sources intended for them would simply not be received.  On the other hand,
if some active users are missed in the activity detection phase, the encoding
scheme proposed in this paper would cause these users to decode to erroneous
source values. To remedy this latter more serious scenario, the BS can 
positively and negatively acknowledge the detected active and inactive users (at a cost of additional downlink bits.)
The design of codes for acknowledgment has been studied in \cite{arq-feedback}. 

\end{remark}
}

\subsection{Achievable Common Message Length}

In this section, we provide upper bounds on the optimal common message length $R^*$ of the
downlink message for massive random access by analyzing the entropy of the encoder output 
$H(g_{\mathbf{m}^*}(\mathbf{X}, \mathbf{A}))$, when
\begin{itemize}
\item Sources ${\mathbf X}=(X_1,\cdots,X_k)$ are distributed according to an exchangeable distribution $p(x_1,\cdots,x_k)$; 
\item Activity pattern ${\mathbf A}$ is uniform among all possible activity patterns and is independent of $\mathbf X$; 
\item The codebook $\mathbf{m}$ is chosen from an ensemble of codebooks 
generated according to i.i.d.\ mixture distribution $q(\mathbf x|\theta)r(\theta)$. 
\end{itemize}

\begin{thm}
\label{thm:main}
Consider a massive random access scenario with a total of $n$ users and a random subset of $k$ active users. Let sources $\mathbf{X} = (X_1, \cdots, X_k)$ take values in a finite or countable set $\mathcal{X}^k$  with exchangeable distribution $p(\mathbf{x})$. The minimum common message length $R^*$ is bounded from above as
\begin{equation}
\label{eq:Theorem1_1}
R^* < \min_{q \in \mathcal{Q}} H(\mathbf{X}) + D(p\|q) + \log(H(\mathbf{X}) + D(p\|q) + 1) + 2, \\
\end{equation}
where $\mathcal{Q}$ is the family of all i.i.d.\ mixture distributions on $\mathcal{X}^k$ 
defined as in \eqref{eq:codebook_generation}.
Further, the minimum common message length $R^*$ is also bounded from above as:
\begin{equation}
\label{eq:Theorem1_2}
R^* < \min_{q \in \mathcal{Q}} H(\mathbf{X}) + D(p\|q) + \log \left( \log \left(\frac{\bar{q}_{\text{max}}}{\bar{q}_{\text{min}}}\right) + 2 \right) + 4,
\end{equation}
where
\begin{equation}
\bar{q}_{\text{max}} = \sup_{\mathbf{x} \in \mathcal{X}^k : p(\mathbf{x})>0} q(\mathbf{x}); \quad 
\bar{q}_{\text{min}} = \inf_{\mathbf{x} \in \mathcal{X}^k : p(\mathbf{x})>0} q(\mathbf{x}).
\label{eq:definition_q_max_min}
\end{equation}
\end{thm}

\begin{IEEEproof}
See Appendix \ref{app:proof_theorem_1}. 
\end{IEEEproof}


It is worth noting that the $D(p\|q)$ term is the divergence between the joint distributions $p(\mathbf{x})$ and $q(\mathbf{x})$, i.e.,
\begin{equation}
    D(p\|q) = \sum_{\mathbf{x} \in \mathcal{X}^k} p(x_1, \cdots, x_k) \log \left( \frac{p(x_1, \cdots, x_k)}{q(x_1, \cdots, x_k)} \right).
\end{equation}
In the setting of this paper, the above divergence term does not necessarily tensorize, as neither the distribution of the sources nor the distribution of the codebook entries are necessarily i.i.d.. However, the operational interpretation of the $D(p\|q)$ term is the same as in classical source coding; it corresponds to the extra cost of using a codebook constructed according to $q({\mathbf x})$ to compress sources with distribution $p({\mathbf x})$.

\begin{cor}
\label{cor:iid}
Under the setting of Theorem \ref{thm:main}, if the sources $\mathbf X$ have an i.i.d.\ mixture distribution $p(\mathbf{x})$, then 
\begin{multline}
R^* < H(\mathbf X) + \min \bigg\{ \log(H(\mathbf X) + 1) + 2 \label{eq:Cor1_1}, \\
	\log \left( \log \left(\frac{\bar{p}_{\text{max}}}{\bar{p}_{\text{min}}}\right) + 2 \right) + 4 \bigg\},
\end{multline}
where 
\begin{equation}
\bar{p}_{\text{max}} = \sup_{\mathbf{x} \in \mathcal{X}^k : p(\mathbf{x})>0} p(\mathbf{x}); \quad 
\bar{p}_{\text{min}} = \inf_{\mathbf{x} \in \mathcal{X}^k : p(\mathbf{x})>0} p(\mathbf{x}).
\label{eq:definition_p_max_min}
\end{equation}
Further, if the sources $\mathbf X$ are i.i.d.\ distributed according to $p(x)$, we have
\begin{multline}
    R^* < k H(p) + \min\bigg\{ \log(k H(p) + 1) + 2, 
\\ \left.  
\log \left( k \log \left(\frac{p_{\text{max}}}{p_{\text{min}}}\right) + 2 \right) + 4\right\}, \label{eq:Cor1_2}
\end{multline}
where 
\begin{equation}
    p_{\text{max}} = \sup_{{x} \in \mathcal{X} : p({x})>0} p({x}); \quad 
    p_{\text{min}} = \inf_{{x} \in \mathcal{X} : p({x})>0} p({x}).
\end{equation}
\end{cor}
\begin{IEEEproof}
The corollary follows directly from Theorem \ref{thm:main} by using a codebook constructed
according to $q(\mathbf x) = p(\mathbf x)$. 
\end{IEEEproof}

Setting $q(\mathbf x) = p(\mathbf x)$ is a natural choice for i.i.d.\ or i.i.d.\
mixture sources, because it maximizes the probability of matching codewords
with the sources. This result shows that for i.i.d.\ and i.i.d.\ mixture sources,
the proposed code construction and the proposed encoding and decoding scheme achieves the joint entropy of the sources to within at most an $O(\log(k))$ overhead. 
The overhead can be further reduced 
if the distribution is close to uniform. In particular, for an i.i.d.\ uniform
distribution with $p_{\text{max}} = p_{\text{min}}$, the entropy can be achieved to within 
$O(1)$ bits. 

An intuitive argument of why for i.i.d.\ sources, a
matching i.i.d.\ codebook can approach the joint entropy of the sources is the following.
When the codebook distribution and the distribution of the sources $p(\mathbf x)$
match, for every realization of the sources $\mathbf x$, a randomly generated
codeword would match $\mathbf x$ with a probability $p(\mathbf x)$. Thus, the
index of the first matching codeword follows a geometric distribution
with parameter $p(\mathbf x)$. The entropy of such a geometric distribution is
essentially $- \log(p(\mathbf x))$ plus a constant. Then, when
averaged over all realizations of $\mathbf x$, we can achieve a common message length of $\mathbb E[- \log(p(\mathbf X))]$,  which is just the entropy of the sources $H(\mathbf X)$, plus a constant. 

To make the above argument work for any codebook distribution $q(\mathbf x)$, which may not be the same as the source distribution $p(\mathbf x)$, we recognize that in general the distribution of the first matching index is a mixture of geometric distributions, for which the exact computation of entropy is complicated.
The proof in Appendix \ref{app:proof_theorem_1} uses a bounding technique due to \cite{sfrl}, 
averaged over randomly generated codebooks, to show the existence of one codebook whose output
has the desired entropy, up to an additive logarithmic term.

\subsection{Coding for Exchangeable Sources}
\label{section:exchangeable-sources}

The proposed codebook construction uses an i.i.d.\ mixture to generate the random
codebook. This is because we do not know in advance which of the $k$ out of 
$n$ users would become active.  Using a mixture of i.i.d.\ distributions to
generate codewords ensures that any $k$ distinct entries of the codewords
all have the same joint distribution. 

If the sources are also i.i.d.\ mixture, then matching the codebook and the distribution of the sources 
allows us to bound the overhead on the common message length to $O(\log(k))$. 
However, for general exchangeable sources which are not i.i.d.\
mixtures, one can no longer simply set the codeword distribution to be the distribution of the sources.
In this section, we develop a novel code construction to allow a generalization of the result of Corollary \ref{cor:iid} from i.i.d.\ sources to sources with arbitrary exchangeable distribution $p({\mathbf x})$.  

One may be tempted to use the marginal distribution $p(x_1)$ of the joint
distribution of the sources $p(x_1,\cdots,x_k)$ to generate i.i.d.\ codewords in
the codebook. But this is not necessarily a good strategy. 
Consider an example where $(X_1,\cdots,X_k)$ takes on only two values,
either $(0,\cdots, 0)$ or $(1,\cdots, 1)$ with probability $0.5$ each. Using the marginal
distribution would have resulted a Ber$(\frac{1}{2})$ codebook for which finding a matching
index would have required searching over $O(2^k)$ codewords, resulting in
a common message length of $O(k)$, whereas the optimal codebook should just have two codewords, i.e., the all-zero and the all-one vectors over the $n$ users,
which would yield a common message length 
of only one bit.

The preceding example motivates us to define the following \emph{urn codebook} for arbitrary
exchangeable sources. Subsequently, we 
define \emph{extended-urn codebook} for exchangeable sources which are extendable.

\subsubsection{Urn Codebook}
The challenge lies in constructing length-$n$ codewords whose arbitrary
subsequences of length $k$ all ``look like'' $\mathbf{x}$.  Recall from
Theorem~\ref{thm:main} that the achievable common message length for communicating
general exchangeable sources $\mathbf{X} = (X_1, \cdots, X_k) \sim p(\mathbf{x})$ is $H(\mathbf{X})$ 
plus an overhead of $D(p\|q)$, where $q$ is an i.i.d.\ mixture 
distribution used for codebook construction. 

The question is then the following. 
For arbitrary exchangeable sources with distribution $p({\mathbf x})$, 
how should we design an i.i.d.\ mixture $q({\mathbf x})$ such that $D(p\|q)$ is small? 
We answer this question by introducing and analyzing the following novel codebook construction. 


\begin{defn}[Urn Codebook]
\label{def:urn-book}
Given exchangeable sources $\mathbf{X}$ with distribution $p({\mathbf x})$, an urn codebook $\mathbf{m}_\text{URN} = \left(\mathbf{c}^{(1)}, \mathbf{c}^{(2)}, \cdots \right)$ is a codebook consisting of codewords $\mathbf{c}^{(j)} \in \mathcal X^n$ generated in the following fashion:
\begin{enumerate}
    \item Generate a realization of $\mathbf{x}=(x_1,\cdots,x_k)$ according to the distribution $p(\mathbf{x})$.
    \item Generate each entry of $\mathbf{c}^{(j)}$ in an i.i.d.\ fashion according to $\hat{p}_\mathbf{x}$, where $\hat{p}_\mathbf{x}$ is the empirical distribution (or the type) of $\mathbf{x}$, i.e., 
    \begin{equation}
    \label{eq:empirical_urn}
	    \hat{p}_\mathbf{x}(c) = \frac{1}{k} \sum_{i=1}^{k} \mathbbm{1}(x_i=c), \quad \forall c \in \mathcal{X}.
    \end{equation}
\end{enumerate}
\end{defn}
It is easy to see that the urn codebook is constructed from a mixture of i.i.d.\ distributions. In particular, the mixture weights are given by the probability that the sampled realization is of a particular type. Further, we argue that this i.i.d mixture is close to the distribution of the sources. To see this, note that this codeword generation process can be alternatively viewed as repeatedly sampling with replacement from an urn containing the elements of $\mathbf{x}$. {\color{black} If instead, every subset of $k$ entries in a codeword can all be viewed as if they are generated by sampling without replacement from the entries of $\mathbf{x}$, then the entries of the codeword would be distributed as $p(\mathbf{x})$. Thus, the difference between the exchangeable source distribution and the i.i.d.\ mixture codebook distribution is precisely the difference between mixtures of sampling with and without replacement.} This intuition is captured in the following theorem, which quantifies this difference by bounding the KL divergence between the distribution of the sources and the i.i.d.\ mixture induced by the code construction.

\begin{thm}
\label{thm:exch}
   Let $p(\mathbf{x})$ be an exchangeable distribution. Let $q(\mathbf{x})$ be the distribution generated by randomly choosing $k$ distinct entries from the codewords in the urn codebook, 
then 
   \begin{align}
	   \label{eq:theorem_2}
	   D(p\|q) \leq \min\left\{\log\left(\frac{k^k}{k!}\right), |\mathcal{X}|\log(k+1)\right\}.
   \end{align}
The first term in the min operation in the above expression is further bounded from above by $k\log(e)$.
\end{thm}
\begin{IEEEproof}
See Appendix \ref{app:theorem_2}.
\end{IEEEproof}

The above theorem gives upper bounds on the cost of using a codebook generated from
an i.i.d.\ mixture distribution to represent sources which are exchangeable. The above
bound immediately leads to the following achievability result for coded
downlink massive random access.

\begin{cor}
\label{cor:exch}
    Consider a massive access scenario with a total of $n$ users and a random subset of $k$ active users. Let sources $\mathbf{X} = (X_1, \cdots, X_k)$ take values in a discrete set $\mathcal{X}^k$ and be distributed according to an exchangeable distribution $p(\mathbf{x})$. Then the minimum common message length $R^*$ for communicating each $X_i$ to its corresponding user $i$ is bounded from above as
    \begin{equation}
    \label{eq:thm-2-1}
        R^* < H(\mathbf{X}) + k\log(e) + \log(H(\mathbf{X}) + k\log(e) + 1) + 2
    \end{equation}
and 
    \begin{multline}
    \label{eq:thm-2-2}
        R^* < H(\mathbf{X}) + |\mathcal{X}|\log(k+1) \\ \qquad\quad+ \log(H(\mathbf{X}) + |\mathcal{X}|\log(k+1) + 1) + 2.
    \end{multline}
\end{cor}
\begin{IEEEproof}
The proof follows directly from Theorem \ref{thm:exch} and Theorem \ref{thm:main}. 
\end{IEEEproof}

The implication is that for exchangeable sources, the overhead beyond $H({\mathbf X})$ is at most $O(k)$, or in the finite alphabet case, at most $O(\log(k))$.

\begin{remark}
It is interesting to note that the $O(k)$ scaling in overhead can already be
obtained using a scheduling code of \cite{scheduling} to order the randomly
activated $k$ users using at most $(k+1)\log(e)$ bits, followed by entropy 
coding of the sources listed in that order, as mentioned earlier.  This approach
of scheduling followed by source coding is different from the strategy of using the urn
codebook, but the two achieve about the same $k \log(e)$ overhead for general exchangeable sources. 
The urn codebook can be thought of as combining scheduling and source coding steps
together. The urn codebook also achieves a smaller overhead in case of finite alphabet sources. 
\end{remark}

\subsubsection{Extended Urn Codebook} 
Although i.i.d.\ and i.i.d.\ mixture sources are also exchangeable, they
incur less than $O(k)$ overhead. The reason is that i.i.d.\ and i.i.d.\
mixtures are \emph{infinitely} extendable. 
In this section, we investigate coding for exchangeable sources that are \emph{finitely} 
extendable, and show that a better common message length can already be achieved in this case.
Recall that sources $(X_1,\cdots,X_k)$ are \emph{$d$-extendable} if they have the
same distribution as the first $k$ elements of a longer exchangeable sequence
$(X_1, \cdots, X_k, \cdots, X_d)$. The idea is that if the sources are $d$-extendable, 
we can construct modified version of the urn codebook by sampling from
realizations of $(X_1,\cdots,X_k,\cdots,X_{d})$. 



\begin{defn}[Extended Urn Codebook]
\label{def:extended-urn-book}
 Given exchangeable sources $(X_1, \cdots, X_k)$ which are $d$-extendable, i.e., there is an extended exchangeable sequence $(X_1, \cdots, X_k, \cdots, X_d)$ with distribution $p(x_1, \cdots, x_d)$, an extended urn codebook $\mathbf{m}_\text{EX-URN} = \left(\mathbf{c}^{(1)}, \mathbf{c}^{(2)}, \cdots \right)$ is a codebook consisting of codewords $\mathbf{c}^{(j)} \in \mathcal X^n$ generated in the following fashion:
 \begin{enumerate}
     \item Sample a realization of $\mathbf{x}=(x_1,\cdots, x_d)$ using distribution $p(x_1, \cdots, x_d)$.
     \item Generate each entry of $\mathbf{c}^{(j)}$ in an i.i.d.\ fashion according to $\hat{p}_\mathbf{x}$, where $\hat{p}_\mathbf{x}$ is the empirical distribution of $\mathbf{x}$:
     \begin{equation}
    \label{eq:empirical_extended_urn}
         \hat{p}_\mathbf{x}(c) = \frac{1}{d} \sum_{i=1}^{d} \mathbbm{1}(x_i=c), \quad \forall c \in \mathcal{X}.
     \end{equation}
 \end{enumerate}
\end{defn}

This is essentially the same definition as Definition \ref{def:urn-book}, but
with sampling occurring on an extended sequence $(X_1, \cdots, X_{d})$ instead.  
Since we are sampling $k$ entries from a larger
vector of size $d$, we expect this distribution to be closer in KL divergence
than the one induced by Definition \ref{def:urn-book}. 

Finite de Finetti theorems are a class of results that provide bounds on 
some measure of distance between an i.i.d.\ mixture distribution (which is how the
codewords in the extended urn codebook are constructed) and an exchangeable and
$d$-extendable distribution. We discuss finite de Finetti results in detail 
in Section \ref{section:finetti}. 
The following is a finite de Finetti theorem under KL divergence. 
It allows a characterization of the minimum common message length for
exchangeable and $d$-extendable sources.

\begin{thm}
\label{thm:exch_extend}
   Let $p(\mathbf{x})$, $\mathbf x \in \mathcal X^k$ be an exchangeable distribution that is also $d$-extendable with $d > k$. 
Let $q(\mathbf{x})$ be the distribution generated by choosing $k$ distinct entries from the codewords in the extended urn codebook, 
then 
   \begin{equation}
	   \label{eq:theorem_3}
        D(p\|q) \leq \min \left\{ 
\log \left(\frac{d^k}{d^{\underline{k}}} \right),
(|\mathcal{X}|-1) \log \left(\frac{d-1}{d-k} \right) \right\}.
   \end{equation}
\end{thm}
\begin{IEEEproof}
See Appendix \ref{app:exch_extend}.
\end{IEEEproof}

\begin{cor}
\label{cor:exch_extend}
    Consider a massive access scenario with a total of $n$ users and a random subset of $k$ active users. Let sources $\mathbf{X} = (X_1, \cdots, X_k)$ take values in a finite or countable set $\mathcal{X}^k$ and be distributed according to an exchangeable and $d$-extendable distribution $p(\mathbf{x})$ with $d > k$. The minimum common message length $R^*$ for communicating each $X_i$ to the respective user $i$ is bounded from above as
    \begin{multline}
    \label{eq:thm-3-1}
        R^* < H(\mathbf{X}) +  \log \left( \frac{d^k}{d^{\underline{k}}} \right)  \\ + \log\left( H(\mathbf{X}) + \log \left( \frac{d^k}{d^{\underline{k}}} \right)  + 1 \right) + 2
    \end{multline}
and 
    \begin{multline}
    \label{eq:thm-3-2}
        R^* < H(\mathbf{X}) + (|\mathcal{X}|-1) \log \left(\frac{d-1}{d-k} \right) \\ \qquad\qquad + \log\left( H(\mathbf{X}) + (|\mathcal{X}|-1) \log \left(\frac{d-1}{d-k} \right) + 1\right) + 2.
    \end{multline}
\end{cor}
\begin{IEEEproof}
The proof follows directly from Theorem \ref{thm:exch_extend} and Theorem \ref{thm:main}. 
\end{IEEEproof}

\begin{remark}
Observe that the first bound in \eqref{eq:theorem_3} essentially reduces to the $k \log(e)$ bound 
in \eqref{eq:theorem_2} in the case of $d=k$. For general exchangeable sources, this bound
is tight. Theorem \ref{thm:definetti_upper} later in the paper shows an example
that achieves this bound. The second terms in
\eqref{eq:theorem_2} and \eqref{eq:theorem_3} are the respective upper bounds 
in the case of finite alphabet sources. These bounds are derived using different proof techniques;
the latter does not reduce to the former in the case of $d=k$. 
\end{remark}

{\color{black}
\begin{remark}
The finite alphabet-size bound in \eqref{eq:theorem_3} and \eqref{eq:thm-3-1}, i.e.,
\begin{equation}
(|\mathcal{X}|-1) \log \left(\frac{d-1}{d-k} \right)
\end{equation} 
is due to the analysis in \cite{matus} on the KL divergence of the probability 
distributions resulting from sampling with and without replacement. 
It is possible to replace this term by 
\begin{equation}
(|\mathcal{X}|-1)\frac{k(k-1)}{2(d-1)(d-k+1)}
\end{equation} 
as in a similar analysis in \cite{stam}.
We explore this connection further in Section \ref{sec:finite_de_finetti}.
\end{remark}
}

\subsection{Applications}
\label{section:examples}

\subsubsection{I.I.D. Sources}
As already shown in Corollary \ref{cor:iid}, when the sources are i.i.d.\ or 
i.i.d.\ mixture distribution, the minimum downlink common message length is
the entropy of the sources, plus an overhead of at most $O(\log(k))$.
This overhead can be reduced to $O(1)$ if the sources are i.i.d.\ and uniformly distributed.

\subsubsection{Scheduling} 
\label{sec:examples-scheduling}
Suppose that we wish to assign the $k$ active users into $b$ slots, where no
two users can be assigned to the same slot.  This is known as the scheduling
problem. Let $X_i \in \{1,\cdots,b\}$ be the slot assignment for the $i$th
user. If all $b \choose k$ schedules are equally likely, then $\mathbf X =
(X_1, \cdots, X_k)$, i.e., the assignment of active users to the slots, is an
exchangeable random sequence. 
We have $H(\mathbf X) = \log \left( {b \choose k} k! \right)$.

Consider the problem of communicating a particular schedule to the $k$ active
users where each user is only interested in learning its own slot assignment.
For the case of $b=k$, the urn codebook would generate codewords according to
uniform distribution on $[k]$. Therefore, we can use \eqref{eq:Theorem1_2} in Theorem \ref{thm:main} to show that the following common message length is achievable 
\begin{eqnarray}
R^* & < & H(\mathbf{X}) + D(p\|q) + 5 \\
& \le & \log(k!) + k \log(e) + 5  \\ 
& \approx & k \log(k),
\label{eq:keqb}
\end{eqnarray}
where $D(p\|q)$ is bounded as in Theorem \ref{thm:exch}.

When $b>k$, this is an example of exchangeable sources with distribution
$p(\mathbf{x})$ that is $b$-extendable, because we can think of the process
of scheduling $k$ users into $b$ slots as the first $k$ steps in the 
scheduling of $b$ users into $b$ slots (which is an extended exchangeable distribution). 

Using the extended urn codebook, Corollary \ref{cor:exch_extend} allows us 
to bound the minimum common message length as
\begin{eqnarray}
R^* & < & H(\mathbf{X}) + D(p\|q) + 5 \\
&\le & \log \left({{b}\choose{k}}k!\right) + \log \left( \frac{b^{k}}{b^{\underline{k}}} \right) + 5 \\
& \approx & k \log(b).
\label{eq:kneqb}
\end{eqnarray}
The achievable common message lengths in \eqref{eq:keqb} and \eqref{eq:kneqb} have an intuitive
explanation, as scheduling a user into one of the $b$ slots requires
$\log(b)$ bits; so the minimum common message length for scheduling $k$ users
should scale as $k \log(b)$. 

A converse for these common message lengths, which is tight to within a constant at large $n$, for the general case of $b \ge k$, can be obtained as follows:
\begin{equation}
\label{eq:converse_scheduling}
    R^* \geq k \log(b) - \log \left( \frac{n^k}{n^{\underline{k}}} \right),
\end{equation}
where $n$ is the total number of users. 
The proof uses a volume bound technique and is presented in Appendix \ref{app:converse_scheduling}.
For any fixed $k$, the second term vanishes
as $n \to \infty$. Therefore, for large $n$, the achievable common message lengths
\eqref{eq:keqb} and \eqref{eq:kneqb} are optimal to within a constant.

In the context of scheduling for massive random access, a different problem is studied
in \cite{scheduling}, where the BS can
choose any schedule for the active users, as long as the slot assignments are
non-colliding.  In this case, assuming $b=k$, the common message length can be
further reduced by a $\log(k!)$ factor, because any of the $k!$ non-colliding
schedules are equally allowable.  This results in a common message length of at most 
$(k+1) \log e$ bits as established in \cite{scheduling} and in agreement with 
the scaling in \eqref{eq:keqb}, where $R^*-\log(k!) \approx k \log e$. 


\subsubsection{Categorization}
\label{sec:examples-categorization}
Consider the task of categorizing the $k$ active users into $c$ categories,
where each category must have a fixed number of $k_\ell$ users, $\ell \in [c]$,
and $\sum_{\ell = 1}^{c} k_{\ell} = k$.  The BS wants to transmit label $\ell
\in [c]$ to the $k_{\ell}$ users in category $\ell$. Each user is only interested
in its own category. 

Assuming that all category assignments satisfying the constraint
$\sum_{\ell = 1}^{c} k_{\ell} = k$ are equally likely, 
the sources $(X_1, \cdots, X_k)$, which represent the category 
labels for the $k$ active users, form an exchangeable random sequence.
This exchangeable random sequence can also be thought of as drawing
$k$ balls from an urn without replacement, where the urn contains $k$
balls in total; the balls are labeled with $\ell \in [c]$; and  
there are $k_\ell$ balls with label $\ell$. 

To communicate the category labels to the active users, we can use 
an urn codebook, where the entries of the codewords can be thought 
of as drawing $k$ balls from the same urn, but with replacement.
Notice that the urn contains $k_\ell$ balls for each label $\ell$ 
with $k_\ell$ fixed, so the codebook distribution $q(\mathbf{x}) = 
\prod_{\ell=1}^{c} \left( \frac{k_\ell}{k} \right)^{k_\ell}$ is a 
constant over the support of $p(\mathbf x)$.  So, we can apply
\eqref{eq:Theorem1_2} in Theorem \ref{thm:main} to show that the following common message length is achievable: 
\begin{equation}
    R^* < H(\mathbf{X}) + D(p\|q) + 5.
\end{equation}
By Theorem \ref{thm:exch}, the overhead $D(p\|q)$ is at most $\min\{k \log(e), c \log(k+1) \}$.

The above achievable common message length can be computed more directly by
expanding out the entropy terms: 
\begin{eqnarray}
    H(\mathbf{X}) + D(p\|q) &=& \sum_{\mathbf{x} \in [c]^k} p(\mathbf{x})\log \left( \frac{1}{q(\mathbf{x})} \right) \\
    &=& \log \left( \frac{1}{\prod_{\ell=1}^{c} \left( \frac{k_\ell}{k} \right)^{k_\ell}} \right) \\
    &=& k \sum_{\ell=1}^{c} \left( \frac{k_\ell}{k} \right) \log \left( \frac{k_\ell}{k} \right)\\ & =& kH(\rho),
\end{eqnarray}
where $\rho = \left( \frac{k_1}{k}, \cdots,  \frac{k_c}{k} \right)$ is the empirical distribution of the labels. Thus interestingly, a characterization of achievable common message length is simply the entropy of the category sizes, i.e., 
\begin{equation}
\label{eq:categorize_rate}
    R^* < kH(\rho) + 5.
\end{equation}
For this specific example, it is also possible to prove a converse that shows 
\begin{equation}
\label{eq:converse_categorization}
    R^* \geq kH(\rho) - \log \left( \frac{n^k}{n^{\underline{k}}} \right),
\end{equation}
where $n$ is the total number of users. The proof is in Appendix
\ref{app:converse_categorization}. Again, for any fixed $k$, the second term vanishes
as $n \to \infty$. Therefore, for large $n$, the achievable common message length 
\eqref{eq:categorize_rate} is optimal to within a constant.

\subsubsection{Resource Allocation}

Suppose that $\mathbf{X} = (X_1,\cdots,X_k)$ must satisfy $\sum_{i=1}^k X_i = \alpha$, where $\alpha \in \mathbb{N}$ and $0 \le X_i \leq \alpha$ takes on integer values. This can be thought of as a resource allocation problem where $\alpha$ units of resources are allocated to $k$ randomly activated users (among a total of $n$ potential users). Consider the joint distribution induced by the process of allocating each unit of resource to the $k$ users uniformly at random one at a time. The resulting $(X_1,\cdots, X_k)$ is exchangeable, but $X_i$'s are not independent due to the constraint $\sum_{i=1}^k X_i = \alpha$. 

The distribution of the sources in this setting is the multinomial distribution:
\begin{equation}
    \mathrm{Pr} (\mathbf{X} = \mathbf{x}) = 
\begin{cases}
    \frac{\alpha!}{x_1!\cdots x_k!}\left(\frac{1}{k}\right)^\alpha, & \text{if } \sum_{i=1}^{k}x_i = \alpha \\
    0, & \text{otherwise}.
\end{cases}
\end{equation}
The multinomial distribution is exchangeable. By Corollary \ref{cor:exch}, 
using the urn codebook achieves a common message length of $H(\mathbf{X})$ bits 
with an overhead of $k\log(e)$ plus a logarithmic term. 

In the previous three examples, the respective urn codebooks all happen to reduce to a construction 
where the codewords are i.i.d.\ generated according to the marginal of the joint distribution. 
But this is not true here. 
We use this example to demonstrate that using the marginal distribution to construct 
the codebook can be inferior to using the urn codebook. 

If we use the marginal distribution to construct the codebook, the codeword distribution 
would have been $q(\mathbf{x}) = \prod_{i=1}^{k}p(x_i)$, where $p(x_i)$ is the marginal of
the multinomial distribution. Then by Theorem \ref{thm:main}, an achievable common message length 
(up to an additional logarithmic term) is
\begin{eqnarray}
R_\mathrm{marginal} & \approx& H(\mathbf{X}) + D(p\|q) \\
    &=& \sum_{x_1, \cdots, x_k} p(\mathbf{x}) \log \left( \frac{1}{\prod_{i=1}^{k}p(x_i)} \right) \\
    &=& k\sum_{x_1} p(x_1) \log \left( \frac{1}{p(x_1)} \right) \\
    &=& kH(X_1). 
\end{eqnarray}
The marginal distribution of the multinomial is a binomial distribution
$p(x_1) = {{\alpha}\choose{x_1}} \left( \frac{1}{k} \right)^{x_1} \left( 1 - \frac{1}{k} \right)^{\alpha-x_1}$.
The entropy of the marginal can be computed as
\begin{equation}
    H(X_1) = \frac{1}{2}\log \left(2\pi e\left( \frac{\alpha}{k} \right) \left( 1 - \frac{1}{k} \right) \
\right) + O\left( \frac{1}{\alpha}\right). 
\end{equation}
On the other hand, if we use an urn codebook, by Corollary \ref{thm:exch} 
the following common message length is achievable (to within an additional logarithmic term): 
\begin{align}
R_\mathrm{urn} & \approx H(\mathbf{X})+k \log e \nonumber \\
    &  = \frac{k-1}{2}\log(2\pi e\alpha) - \frac{k}{2}\log(k) + O\left( \frac{1}{\alpha}\right) + k \log e, 
\end{align}
where the entropy of the multinomial distribution is computed according to the approximation in \cite{cichon}.

Comparing $R_\mathrm{marginal}$ with $R_\mathrm{urn}$, we see that their difference can be lower bounded as follows:
\begin{align}
&R_\mathrm{marginal} - R_\mathrm{urn} \nonumber \\
    &\approx \frac{1}{2}\log(2\pi e\alpha) + \frac{k}{2}\log\left( 1 - \frac{1}{k} \right) - k \log e + O\left( \frac{1}{\alpha}\right) \\
    &\geq \frac{1}{2}\log(2\pi e\alpha) - \frac{\log(e)}{2} - k \log e + O\left( \frac{1}{\alpha}\right) \\
    &= \frac{1}{2}\log(2\pi \alpha) - k \log e + O\left( \frac{1}{\alpha}\right).
\label{eq:marginal_urn}
\end{align}
Therefore at fixed $k$, for large values of $\alpha$, it is advantageous to use the urn codebook.

The value of $\alpha$ for which the urn codebook outperforms a codebook constructed from the marginal distribution does not need to be as large as what \eqref{eq:marginal_urn} suggests.
 Consider the scenario where $\alpha = k = 2$. In this case, the entropy of the marginal can be computed directly as
\begin{equation}
    H(X_1) = -\sum_{i=0}^{2} p_i \log p_i = 1.5,
\end{equation}
where $p_i = {{2}\choose{i}} \left( \frac{1}{2} \right)^2$. 
Therefore, $kH(X_1) = 3$. 

On the other hand, the entropy of $H(X_1, X_2) = H(X_1) = 1.5$, since the value of $X_2$ is determined by the value of $X_1$. This gap is therefore $1.5$ bits. Now, the extra $k \log(e)$ cost comes from approximating the divergence between the multinomial distribution and the corresponding urn codebook. For finite $k$, the extra cost is bounded from above by 
$\log \left( \frac{k^k}{k!} \right) = 1 < 1.5$ for $k=2$. 
Therefore, the urn codebook is already better than the codebook constructed from the marginal for $\alpha=k=2$. 

The urn codebook in this case contains only two types of codewords: $(1,\cdots,1)$ codeword and codewords with $0$ and $2$ in 50\% proportion each, whereas the codebook constructed from marginal distribution has codewords with mixed 0, 1, and 2 entries, which is not as effective.

\subsection{Random Number of Active Users}
{\color{black}
So far in this paper, we have assumed that the number of active users $k$ is fixed and known
in advance. This is not always realistic, because user activities are typically random
in a massive access scenario. In this section, we consider a scenario in which we wish to transmit sources $(X_1, \cdots, X_K)$ to a set of \emph{randomly} activated users. Let $K$ be a random variable which 
denotes the number of active users. The BS learns the activity pattern, thus the realization of $K$
in each transmission slot, then aims to transmit the sources $(X_1, \cdots, X_K)$ to each of the 
$K$ active users, respectively. 

For the problem of communicating general exchangeable sources of variable size $K$, 
because most random vectors do not have well defined distributions when the size of 
the vector can vary, devising a general coding strategy for such sources is challenging.
However, what one can do is to construct a library of codebooks, one for each 
possible value of $K$. After the set of active users are detected, the BS first 
broadcasts the actual number of active users using $H(K)$ bits, then uses the
codebook designed for the sources $(X_1, \cdots, X_K)$ to communicate to the
$K$ users.  In this way, the total expected common message length is simply the
average common message length averaged over $K$ plus $H(K)$ bits.

However, in the special case that the sources are i.i.d., it turns out
that it is not necessary to communicate the number of active users,
thus avoiding the $H(K)$ bits overhead. The following result is a counterpart of
Corollary \ref{cor:iid} for random number of active users, which shows that the
minimum common message length is essentially the \emph{expected} entropy of the
sources.} 


\begin{thm} 
\label{thm:random_users}
Fix the total number of users $n$. Let the number of active users
$K \in [n]$ be a random variable.  Let $a_i$ be the index of the $i$-th active
user, $i \in [K]$.  Let $(X_1, \cdots, X_K)$ be i.i.d.\ $\sim p(x)$. 
The minimum common message length $R^{*}$ for communicating each $X_i$ to its respective intended user $a_i$, $i = 1, \cdots, K$, is bounded from above as
\begin{equation}
  R^{*} < \mathbb{E}[K]H(p) + \log(\mathbb{E}[K]H(p) + 1) + 2.
\end{equation}
\end{thm}

\begin{IEEEproof}
The proposed coding strategy of using a codebook consisting of codewords of size $n$, then looking for a matching codeword for the active users, works equally well for random $K$ as for fixed $k$. To bound the entropy of matching index, it is possible to first condition on $k$, then average over $K$. The details are in Appendix \ref{app:random_users}.
\end{IEEEproof}

\section{Finite de Finetti Theorem}
\label{section:finetti}

In this section, we discuss the connection between the codebook construction for downlink massive random access and 
the study of finite exchangeable sequences and the finite de Finetti theorems. 

\subsection{Existing Finite de Finetti Theorems}

The original de Finetti theorem states that every infinitely extendable sequence of exchangeable random variables has a distribution that is equivalent to an i.i.d.\ mixture. This result can be generalized to finite exchangeable sequences in what is known as the finite de Finetti theorem. Specifically, it is shown in \cite{dia-fre} that if $(X_1, \cdots, X_k, \cdots, X_d)$ is an exchangeable sequence of random variables, then the distribution of $(X_1, \cdots, X_k) \sim p(\mathbf{x})$ is close to being an i.i.d.\ mixture in the sense that  
\begin{equation}
\label{eq:dia-fre-finetti}
    \min_{q \in \mathcal{Q}} \mathrm{TV}(p, q) \leq \min \left\{  \frac{k(k-1)}{2d},\frac{|\mathcal{X}|k}{d} \right\},
\end{equation}
where $\mathrm{TV}(p,q) = \frac{1}{2}\sum_{\mathbf{x} = \mathcal{X}^k} | p(\mathbf{x}) - q(\mathbf{x}) |$ and $\mathcal{Q}$ is the family of i.i.d.\ mixture distributions on $\mathcal{X}^k$. This is an upper bound on how close finite exchangeable sequences are to being i.i.d.\ mixture in terms of TV distance.

An interesting implication of \eqref{eq:dia-fre-finetti} is that the TV distance between the distribution of a $d$-extendable exchangeable sequence $(X_1, \cdots, X_k)$ and the closest i.i.d.\ mixture distribution can go to $0$ even when $k \to \infty$, if $d$ also goes to infinity in certain ways. In the case of finite alphabet size, i.e., $|\mathcal{X}| < \infty$, the TV distance goes to $0$ as long as $\frac{k}{d} \to 0$, and for general alphabets as long as $\frac{k^2}{d} \to 0$. Examples are presented in \cite{dia-fre} that show these scalings cannot be improved upon in general.

There has been a recent interest in establishing similar results as \eqref{eq:dia-fre-finetti}, but in terms of KL divergence. Given that $p(\mathbf{x})$ is the distribution of a $d$-extendable exchangeable sequence $\mathbf{X} = (X_1, \cdots, X_k)$, the aim is to upper bound 
\begin{equation}
\label{eq:min-divergence}
    \min_{q \in \mathcal{Q}} D(p\|q),
\end{equation}
where $\mathcal{Q}$ is the family of i.i.d.\ mixture distributions on $\mathcal{X}^k$.


{\color{black}
Two upper bounds on \eqref{eq:min-divergence} have recently been developed in \cite{exch-it-3} and \cite{exch-it-4}. The first bound is independent of the alphabet size, but depends on the source distribution. It states that
\begin{equation}
\label{eq:ub-with-entropy-2}
    \min_{q \in \mathcal{Q}} D(p\|q) \leq \frac{k(k-1)}{2(d-k+1)} H(X_1).
\end{equation}
The second bound, which depends on the alphabet size, states that
\begin{equation}
\label{eq:ub-using-stam}
    \min_{q \in \mathcal{Q}} D(p\|q) \leq (|\mathcal{X}|-1)\frac{k(k-1)}{2(d-1)(d-k+1)}.
\end{equation}

For the important special case where the exchangeable sequence is not extendable, i.e., $d = k$, the two upper bounds taken together imply that
\begin{equation}
\label{eq:ub-with-d_equal_k}
   \min_{q \in \mathcal{Q}} D(p\|q) \leq \min \left\{ \frac{k(k-1) H(X_1)}{2},\frac{(|\mathcal{X}|-1)k}{2} \right\}.
\end{equation}
We see that the first argument of the min operation depends on $k$ as $O(k^2)$, while the second argument depends on $|\mathcal{X}|$ and $k$ as $O(|\mathcal{X}|k)$. In this paper, we show that both of these scalings can be improved. 

For the case of $d>k$, \eqref{eq:ub-with-entropy-2} and \eqref{eq:ub-using-stam} imply that  
the KL divergence between an exchangeable sequence and its best i.i.d.\ mixture approximation goes to $0$ as long as $\frac{k}{d} \to 0$ for the finite alphabet size case,
and for general alphabets as long as $\frac{k^2}{d} \to 0$. These are the same scalings as implied by \eqref{eq:dia-fre-finetti} for TV distance. 

These scalings are 
optimal not only for TV distance but also for KL divergence. This is because 
although the TV distance cannot be used to upper bound KL divergence, the reverse is possible through Pinkser's inequality \cite{pinsker}. 
Pinsker's inequality states that}
\begin{equation}
\label{eq:pinsker}
    \mathrm{TV}(p,q) \leq \sqrt{\frac{1}{2\log(e)}D(p\|q)}.
\end{equation}
Since the scalings in \eqref{eq:dia-fre-finetti} are the best possible for TV distance for general exchangeable sequences, we have a natural converse for upper bounding the KL divergence between the i.i.d.\ mixtures and the exchangeable and extendable sequences, i.e. the bounds cannot scale better than \eqref{eq:dia-fre-finetti}.

Although the scalings of upper bounds \eqref{eq:ub-with-entropy-2} and \eqref{eq:ub-using-stam} match the optimal scalings of \cite{dia-fre}, the bound \eqref{eq:ub-with-entropy-2} is distribution dependent as it contains the entropy term $H(X_1)$. 
Using the approach developed in this paper, we show that it is possible to remove this distribution dependency.

\subsection{An Improved Finite de Finetti Theorem}
\label{sec:finite_de_finetti}

In the context of downlink massive random access, 
the minimum divergence \eqref{eq:min-divergence} is essentially the overhead for 
the minimum achievable common message length beyond the joint entropy $H(\mathbf{X})$,
according to Theorem \ref{thm:main}. 
For this reason, the characterizations of the common message length for
the urn codebook construction in Theorem \ref{thm:exch} and for the extended urn codebook construction in Theorem \ref{thm:exch_extend} immediately give us the following finite de Finetti theorem based on KL divergence.

{\color{black}
\begin{thm}[Finite de Finetti with KL Divergence]
\label{thm:definetti}
Let $(X_1, \cdots, X_k, \cdots, X_d)$ be a sequence of exchangeable random variables taking values in $\mathcal{X}^d$, where $\mathcal{X}$ is a finite or countable alphabet. Let $p(x_1, \cdots, x_k)$ denote the distribution of $(X_1, \cdots, X_k)$. The minimum divergence is bounded as
\begin{equation}
\label{eq:de-finetti-1}
   \min_{q \in \mathcal{Q}} D(p\|q) \leq  \log \left(\frac{d^k}{d^{\underline{k}}}\right).
\end{equation}
For the case of $d=k$, the right-hand side of the above is further upper bounded as $\log\left( \frac{k^k}{k!} \right) < k\log(e)$. 

When the alphabet size is finite, we also have that
\begin{equation}
\label{eq:de-finetti-2}
\min_{q \in \mathcal{Q}} D(p\|q) \leq  
\begin{cases}
	|\mathcal{X}| \log(k+1) & \mathrm{if}\ d=k \\ 
        \displaystyle (|\mathcal{X}|-1) \log \left(\frac{d-1}{d-k} \right) & \mathrm{if}\ d>k. 
\end{cases}
\end{equation}
\end{thm}}
\begin{IEEEproof}
    The proof follows directly from Theorem \ref{thm:exch} and Theorem \ref{thm:exch_extend}. 
\end{IEEEproof}

{\color{black}

For the special case where the exchangeable sequence is not extendable, i.e., $d=k$, this finite de Finetti bounds is a consequence of Theorem \ref{thm:exch}, i.e., 
\begin{equation}
\label{eq:de-finetti-d=k}
\min_{q \in \mathcal{Q}} D(p\|q) \leq  
\min \left\{ k\log(e),|\mathcal{X}| \log(k+1) \right\}.
\end{equation}
Comparing to \eqref{eq:ub-with-d_equal_k}, we see that the new bound
improves the scaling for the finite alphabet size case 
from $O(|\mathcal{X}|k)$ to $O(|\mathcal{X}|\log(k))$, and for the general alphabet case from $O(k^2)$ to $O(k)$. Furthermore, the first term of \eqref{eq:de-finetti-d=k} is distribution independent, in contrast to the first term in \eqref{eq:ub-with-d_equal_k} which depends on the entropy $H(X_1)$.

For the case of $d>k$, we first analyze the scaling of \eqref{eq:de-finetti-1}. By using the following useful inequality from \cite{rem-samp}, which states that for $d > \frac{1}{2}k(k-1)$,
\begin{equation}
    \log\left(\frac{d^k}{d^{\underline{k}}}\right) \leq -\log \left(1 - \frac{k(k-1)}{2d} \right),
\label{eq:dk_bound}
\end{equation}
we see that the scaling of $d$ and $k$ required to drive the minimum $D(p \| q)$ to zero in the bound \eqref{eq:de-finetti-1} is that $\frac{k^2}{d} \to 0$. For the finite alphabet size case, the scaling required in the bound \eqref{eq:de-finetti-2} is that $\frac{k}{d} \to 0$. 
These scaling results match 
that of \eqref{eq:ub-with-entropy-2} and \eqref{eq:ub-using-stam},
and also that of \eqref{eq:dia-fre-finetti}.
These are therefore the best possible scaling for reasons due to Pinsker's inequality, as mentioned before. 

A key feature of the bound \eqref{eq:de-finetti-1} as compared to \eqref{eq:ub-with-entropy-2} is that it is distribution independent. 
On the other hand, the second case (for $d>k$) in \eqref{eq:de-finetti-2} can be larger or smaller than \eqref{eq:ub-using-stam}, as analyzed in \cite{matus}.
 The second case in \eqref{eq:de-finetti-2} is a result of \cite[Theorem 4.5]{matus}, which bounds the divergence between sampling with and without replacement from an urn containing elements from set $\mathcal{X}$. The bound \eqref{eq:ub-using-stam} is derived using a similar argument, but uses an alternative bound from \cite{stam} instead of \cite[Theorem 4.5]{matus}. A detailed comparison of these two bounds can be found in \cite{matus}.

}

We conclude this section by pointing out that the bound \eqref{eq:de-finetti-1}
is not only scaling optimal, but also
the best possible upper bound for general exchangeable and $d$-extendable sequences. 
We establish this fact by showing that for a particular exchangeable and $d$-extendable
distribution $r$, the above upper bound is achieved. 

\begin{thm}
\label{thm:definetti_upper}
    Fix $d \ge k$. Let $r(x_1, \cdots, x_k)$ be the distribution of $k$ draws without replacement from an urn containing elements $\{ 1, 2, \cdots d \}$, i.e., $|\mathcal X|=d$. Then, for every i.i.d.\ mixture distribution $q(x_1, \cdots, x_k)$, we have that
    \begin{equation}
\label{eq:definetti_lower}
        D(r\|q) \geq \log \left( \frac{d^k}{d^{\underline{k}}} \right).
    \end{equation}
\end{thm}

\begin{IEEEproof}
The proof takes inspiration from \cite{dia-fre} and is deferred to Appendix \ref{app:definitti_upper}. 
\end{IEEEproof}

Thus, for this particular $r(x_1,\cdots,x_k)$, defined when $d \geq k$, the minimum $D(r\|q)$ over $q \in \mathcal{Q}$
is precisely the right-hand side of \eqref{eq:de-finetti-1} and \eqref{eq:definetti_lower}. In the case of $d=k$, this lower bound is $\log\left(\frac{k^k}{k!} \right) \approx k \log e$.

Note that in this example, we have $|\mathcal X|=d$, for which the right-hand side of \eqref{eq:de-finetti-1} is always larger than the right-hand side of \eqref{eq:de-finetti-2}.
For general exchangeable and $d$-extendable $p(x_1,\cdots,x_k)$, e.g., with smaller $|\mathcal X|$, the minimum $D(p\| q)$ may be smaller than the right-hand side of \eqref{eq:de-finetti-1}.
 
\section{Conclusions}
\label{sec:conclusion}

This paper connects the finite de Finetti theorem and the
problem of communicating sources with certain symmetry to a randomly
activated subset of users in downlink massive random access.  We develop a
KL divergence version of the finite de Finetti theorem and provide a
bound on the extra cost of using a codebook generated from an i.i.d.\ mixture 
distribution to encode and send exchangeable sources to a random subset of $k$ users out of a large pool of $n$ users.
The extra cost is independent of $n$, and is at most $O(k)$ for general 
exchangeable sources, $O(\log k)$ for finite alphabet exchangeable sources,
and $O(1)$ for uniform i.i.d.\ sources. 


\section{Acknowledgment}
We thank the anonymous reviewers for their comments which have improved the presentation of this paper and simplified the proof of Theorem 1. 

\appendices 

\section{Proof of Theorem \ref{thm:main}} 
\label{app:proof_theorem_1}


First of all, in a massive random access scenario where a random subset of $k$ users become 
active among a large pool of $n$ users, when the sources are exchangeable,
it does not matter which subset of $k$ users are activated. So we only need to consider
$p(\mathbf{x})$ with $\mathbf x$ taking values on $\mathcal{X}^k$. 

Instead of constructing a specific codebook, 
the main idea of the proof is to bound the entropy of the encoder output over \emph{an ensemble of random codebooks}. 
Let $p(\mathbf{x})$ be an exchangeable distribution on $\mathcal{X}^k$. 
Let $q(\mathbf{x})$ be an i.i.d.\ mixture distribution on $\mathcal{X}^k$.
Define $T = g_\mathbf{M}(\mathbf{X}, \mathbf{A})$ to be the encoder output from
the following process. We generate sources $\mathbf x$ according to $p(\mathbf x)$. 
We also generate an infinite-size random codebook $\mathbf{M}$ according to $q(\mathbf x)$. 
We find $T$, the index of the first codeword in $\mathbf M$ that matches the sources $\mathbf x$, according to \eqref{eq:encoder}. 
We will find an upper bound on $H(T)$. 

If the encoder and the decoders share unlimited common randomness, the upper bound on $H(T)$ would have already given us the desired achievability result.
In case that common randomness is not available, the encoder and the decoders must share a particular infinite-size codebook. 
In this case, based on the upper bound on $H(T)$, we can argue that there must exist
at least one good codebook, whose output entropy across all input $\mathbf x \sim p(\mathbf x)$ is bounded by the same upper bound. 
This is because mixing increases entropy, so 
\begin{equation}
	\overline{H(g_\mathbf{m}(\mathbf{X}, \mathbf{A}))} \leq H(g_\mathbf{M}(\mathbf{X}, \mathbf{A})) = H(T),
\end{equation}
where $H(g_\mathbf{m}(\mathbf{X}))$ is the encoder output entropy for 
a particular codebook $\mathbf m$, and the overline denotes the averaging operation over $\mathbf m$. 
Since there must exist at least one codebook $\mathbf{m}^*$ such that 
\begin{equation}
	H(g_{\mathbf{m}^*}(\mathbf{X}, \mathbf{A})) \leq \overline{ H(g_\mathbf{m}(\mathbf{X}, \mathbf{A})) },
\end{equation}
(as otherwise the average cannot be achieved), it follows that
\begin{equation}
\label{eq:averaging-arg}
    R^* < H(g_{\mathbf{m}^*}(\mathbf{X}, \mathbf{A})) + 1 \leq H(T) + 1.
\end{equation}
Thus in the rest of the proof, we only need to bound $H(T)$.

Next, we analyze the distribution of $T$. 
Since the codewords of a random codebook are 
generated in an i.i.d.\ fashion according to the distribution $q(\mathbf{x})$, it follows 
that when conditioned on the sources $\mathbf{x}$, the probability that the 
first match occurs at $T=t$ is a geometric distribution with parameter $q(\mathbf{x})$.
Then, it follows that the overall distribution of $T$, across all the sources $\mathbf x$, must be a mixture of geometric distributions:
\begin{equation}
\label{eq:geo_mix}
    \mathrm{Pr}(T = t) = \sum_{\mathbf{x} \in \mathcal{X}^k} p(\mathbf{x})(1-q(\mathbf{x}))^{t-1}q(\mathbf{x}).
\end{equation}
We proceed with bounding the entropy of this geometric mixture.

\subsection{Proof of (\ref{eq:Theorem1_1})} 

An exact calculation of the entropy of a mixture of geometric distributions is laborious. 
To circumvent this, we upper bound $H(T)$ using the fact that    
\begin{equation}
\label{eq:max-entropy-arg}
    H(T) \leq \mathbb{E}[\log(T)] + \log(\mathbb{E}[\log(T)] + 1) + 1.
\end{equation}
A proof of \eqref{eq:max-entropy-arg} can be found in \cite{sfrl}, where a maximum entropy argument is used. The problem is now reduced to bounding $\mathbb{E}[\log(T)]$. 

{\color{black}
To do so, we apply the conditional version of Jensen's inequality:
\begin{eqnarray}
    \mathbb{E}[\log(T)|\mathbf{X} = \mathbf{x}] &\leq& \log \left( \mathbb{E}[T| \mathbf{X} = \mathbf{x} ]\right) \\
    &=& \log \left( \frac{1}{q(\mathbf{x})} \right),
\end{eqnarray}
since for a fixed $\mathbf{X} = \mathbf{x}$, the distribution of $T$ is a geometric distribution with success probability $q(\mathbf{x})$. Averaging over $\mathbf{X}$, we have
\begin{eqnarray}
    \mathbb{E}[\log(T)] &=& \sum_{\mathbf{x} \in \mathcal{X}^k} p(\mathbf{x})\mathbb{E}[\log(T)|\mathbf{X} = \mathbf{x}] \\
    &\leq& \sum_{\mathbf{x} \in \mathcal{X}^k} p(\mathbf{x}) \log \left( \frac{1}{q(\mathbf{x})} \right) \\
    &=& H(\mathbf{X}) + D(p\|q).
\end{eqnarray}
Combining with \eqref{eq:max-entropy-arg} and \eqref{eq:averaging-arg}, we have 
\begin{equation}
    R^* < H(\mathbf{X}) + D(p\|q) + \log(H(\mathbf{X}) + D(p\|q)+1)+2.
\end{equation}
An alternative proof for the above can be obtained by direct computation; see \cite{coded-downlink}.
}

\subsection{Proof of (\ref{eq:Theorem1_2})} 

We bound the entropy of $T$ by defining a new random variable $L$ and using the fact that
\begin{equation}
\label{eq:basic-entropy-bound}
    H(T) \leq H(T,L) = H(L) + H(T|L). 
\end{equation}
The idea here is to choose an $L$ such that the right-hand side of the above expression is easy to analyze. 
To this end, we define $L$ based on the following partition $\mathcal{D}_1, \mathcal{D}_2 \ldots \subseteq \mathcal{X}^k$ where
\begin{equation}
    \mathcal{D}_\ell = \left\{ \mathbf{x} \in \mathcal{X}^k \ \bigg| \ \frac{1}{2^{\ell}} \leq q(\mathbf{x}) < \frac{1}{2^{\ell-1}}, \ p(\mathbf{x})>0 \right\}.
\end{equation}
Assuming that $q(\mathbf{x})$ is non-degenerate, the sets $\mathcal{D}_\ell$ can be thought of as a categorization of all possible sources based on their probability of occurrence with respect to $q(\mathbf{x})$, i.e., the distribution used to generate the codebook. Since $\mathcal{D}_1, \mathcal{D}_2 \ldots$ form a partition, each realization of $\mathbf{x}$ is contained in exactly one of these sets. We can define $L(\mathbf x)$ as the index of the set containing $\mathbf{x}$.
\begin{equation}
    L(\mathbf x) = \ell, \quad \text{s.t.} \ \ \mathbf{x}  \in \mathcal{D}_\ell. 
\end{equation}
It follows that for any $\mathbf{x}\in\mathcal{X}^k$ with $p(\mathbf{x})>0$, the index of the set that $\mathbf{x}$ belongs to is $\ell = \left\lceil \log \left( \frac{1}{q(\mathbf{x})} \right) \right\rceil$. This means that $L$ takes on at most {\color{black}$\left\lceil \log \left( \frac{1}{\bar{q}_\text{min}} \right) \right\rceil - \left\lceil \log \left( \frac{1}{\bar{q}_\text{max}} \right) \right\rceil + 1$} values (infinite if $\bar{q}_{\text{min}} = 0$), and therefore has an entropy bounded from above as
\begin{align}
\label{eq:binning-z}
    H(L) &\leq \log \left( \left\lceil \log \left( \frac{1}{\bar{q}_{\text{min}}} \right) \right\rceil - \left\lceil \log \left( \frac{1}{\bar{q}_{\text{max}}} \right) \right\rceil + 1 \right) \\
         &\leq \log\left(\log\left(\frac{\bar{q}_{\text{max}}}{\bar{q}_{\text{min}}}\right)+2\right).
         \end{align}
It remains to bound $H(T|L)$ in~\eqref{eq:basic-entropy-bound}. 
Given $L=\ell$, we know that the sources must be in $\mathcal{D}_\ell$ and therefore we can compute the conditional probability mass function
\begin{equation}
    \mathrm{Pr}(T=t|L=\ell) = \sum_{\mathbf{x}\in \mathcal{D}_\ell}\frac{p(\mathbf{x})}{\mathrm{Pr}(L=\ell)}(1-q(\mathbf{x}))^{t-1}q(\mathbf{x}).
\end{equation}
We can see that $L$ acts as a way to quantize the original mixture distribution. Letting $r_\ell(t) = \mathrm{Pr}(T=t|L=\ell)$, we can upper bound $H(T|L=\ell)$ as
\begin{align}
\label{eq:cond-entropy-ub}
    H(T|L=\ell) &= -\sum_{t=1}^{\infty} 
    r_\ell(t)\log \left({r_\ell(t)} \right) \\
    &\leq -\sum_{t=1}^{\infty}r_\ell(t)\log \left( {\mathcal G(t;2^{-\ell})} \right),
 \end{align}
where $\mathcal G(t;\theta)$ is a geometric distribution with parameter $\theta$ and the second line follows from the fact that 
$- \sum_{t}r_\ell(t)\log \left( r_\ell(t)\right) \leq - \sum_{t}r_\ell(t)\log \left( r'(t)\right)$  for any distribution $r'(t)$.  We can now bound $H(T|L)$  using \eqref{eq:cond-entropy-ub} 
{\color{black}
\begin{align}
\label{eq:binning-v-z2}
H(T|L) &= \sum_{\ell=1}^{\infty}\mathrm{Pr}(L=\ell)H(T|L=\ell) \\
&\leq  -\sum_{\ell=1}^{\infty} \sum_{\mathbf{x}\in \mathcal{D}_\ell} p(\mathbf{x}) \sum_{t=1}^{\infty}r_\ell(t)\log \left( {\mathcal G(t;2^{-\ell})} \right) \\
&= -\sum_{\ell=1}^{\infty} \sum_{\mathbf{x}\in \mathcal{D}_\ell} p(\mathbf{x}) \sum_{t=1}^{\infty}r_\ell(t)\log \left( r_\ell(t) \right) \nonumber\\
&\qquad + \sum_{\ell=1}^{\infty} \sum_{\mathbf{x}\in \mathcal{D}_\ell} p(\mathbf{x}) \sum_{t=1}^{\infty}r_\ell(t)\log \left( \frac{r_\ell(t)}{\mathcal G(t;2^{-\ell})} \right) \\
&= -\sum_{\mathbf{x}} p(\mathbf{x}) \sum_{t=1}^{\infty}\mathcal G(t;q(\mathbf{x}))\log \left( {\mathcal G(t;q(\mathbf{x}))} \right) \nonumber\\
&\qquad + \sum_{\ell=1}^{\infty} \sum_{\mathbf{x}\in \mathcal{D}_\ell} p(\mathbf{x})D(\mathcal G(t;q(\mathbf{x}))\|\mathcal G(t;2^{-\ell})). \label{eq:binning-v-z2-1}
\end{align}}
Note that right-hand side of the expression of~\eqref{eq:binning-v-z2-1} is the sum of two terms. The first is a weighted sum of the entropy terms of geometric random variables, whereas the second is a weighted sum of KL divergences between geometric distributions. Both terms can be upper bounded by using properties of geometric distributions. Using the fact that the entropy of a random variable with distribution $\mathcal G(t;\theta)$ is upper bounded by $\log(1/\theta) + \log(e)$ bits, the first term is upper bounded by $\log(1/q(\mathbf{x})) + \log(e)$. For the second term, it can be shown that $D(\mathcal G(t;q(\mathbf{x}))\|\mathcal G(t;2^{-\ell})) \leq 1$, for $2^{-\ell}\leq q(\mathbf{x}) \leq 2^{-\ell + 1}$. Hence, it follows that
\begin{equation}
    \sum_{\ell=1}^{\infty} \sum_{\mathbf{x}\in \mathcal{D}_\ell} p(\mathbf{x})D(\mathcal G(t;q(\mathbf{x}))\|\mathcal G(t;2^{-\ell})) \leq 1.
\end{equation}
Putting the bounds together, we have
{\color{black}
\begin{align}
\label{eq:binning-v-z}
    H(T|L) &\leq \sum_{\mathbf{x} \in \mathcal{X}^k} p(\mathbf{x}) \left( \log \left( \frac{1}{q(\mathbf{x})} \right) + \log(e) \right) + 1 \nonumber \\
    &< H(\mathbf{X}) + D(p\|q) + 3.
\end{align}}
Finally, combining \eqref{eq:averaging-arg}, \eqref{eq:basic-entropy-bound}, \eqref{eq:binning-z}, and \eqref{eq:binning-v-z} yields 
\begin{equation}
    R^* < H(\mathbf{X}) + D(p\|q) + \log\left( \log\left(\frac{\bar{q}_{\text{max}}}{\bar{q}_{\text{min}}}\right)+2\right) + 4.
\end{equation}

\section{Proof of Theorem \ref{thm:exch}}
\label{app:theorem_2}

As in the proof of Theorem \ref{thm:main}, we only need to consider sources
$\mathbf{x} \in \mathcal{X}^k$ corresponding to any random subset of $k$ active users
among the $n$ users. 

In the urn codebook construction, we first generate a vector $\mathbf{s} \in \mathcal{X}^k$
according to $p(\mathbf{\mathbf{x}})$, then sample the entries of this vector to generate 
codewords $\mathbf{c} \in \mathcal{X}^n$. Note that here we use slightly different notation 
as in the text and call this vector $\mathbf{s}$. 

{\color{black}
Our goal is to characterize $q(\mathbf{x})$, the probability that $\mathbf{x}$ 
appears in any set of $k$ distinct entries in any codeword in the urn codebook. 
We represent the sampling process by defining the positions of $\mathbf{s}$ that 
are sampled as $\mathbf{w} = (w_1, \cdots, w_k)$ where each $w_i$ is i.i.d.\ over $[k]$.
In other words,
\begin{equation}
[x_1, \cdots, x_k]^T = [s_{w_1}, \cdots, s_{w_k}]^T.
\label{eq:sampling_processing}
\end{equation}
In the rest of the proof, we use the shorthand 
``$\mathbf{x}=\mathbf{s}_\mathbf{w}$''
to denote the above. 

Now, if the sampling pattern were collision-free, it would have
generated sequences that are distributed according to $p({\mathbf x})$ exactly. 
Define the set of all collision-free $\mathbf{w}$ as
\begin{equation}
\mathcal{F} = \{ \mathbf{w} \in [k]^k \ | \ w_i \neq w_j\ \forall i \neq j \}.
\end{equation}

The actual sampling process may not be collision-free, but we can bound the distribution $q(\mathbf{x})$ as follows: 
\begin{align}
    q(\mathbf{x}) &= 
\sum_{\mathbf{s} \in \mathcal{X}^k} 
\sum_{\mathbf{w} \in [k]^k} 
p(\mathbf{s}) 
p(\mathbf{w}) 
\mathbbm{1}({\mathbf{x}=\mathbf{s}_\mathbf{w}}) \label{eq:counting_1} \\ 
    &\geq 
\sum_{\mathbf{s} \in \mathcal{X}^k} 
\sum_{\mathbf{w} \in \mathcal{F}}
p(\mathbf{s}) 
p(\mathbf{w}) 
\mathbbm{1}({\mathbf{x}=\mathbf{s}_\mathbf{w}}) \label{eq:counting_2} \\ 
    &= \sum_{\mathbf{w} \in \mathcal{F}} \frac{1}{k^k}p(\mathbf{x}) \label{eq:counting_3} \\
    & = \frac{k!}{k^k}p(\mathbf{x}). \label{eq:counting_4}
\end{align}
where 
\begin{itemize}
\item \eqref{eq:counting_1} is due to the urn codebook construction process;
\item \eqref{eq:counting_2} is due to that collision-free sampling is a subset of all samplings; 
\item \eqref{eq:counting_3} is due to the facts that under collision-free sampling $p(\mathbf{x}) = p(\mathbf{s})$ and that $p(\mathbf{w}) = \frac{1}{k^k}$ since all sampling patterns are equally likely; 
\item \eqref{eq:counting_4} is due to the fact that the number of collision-free samplings is  $| \mathcal{F} | = k!$. 
\end{itemize}

Noting that $k! > k^ke^{-k}$, we have $\log\left(\frac{k^k}{k!} \right) < k\log(e)$. This implies that 
\begin{eqnarray}
D(p\|q) & \le & \sum_{\mathbf{x} \in \mathcal{X}} p(\mathbf{x}) \log \left( \frac{p(\mathbf{x})}{q(\mathbf{x})}\right) \\ 
& \leq & \log\left(\frac{k^k}{k!} \right) \label{eq:theorem_2_1_2}\\
& < & k\log(e). 
\label{eq:theorem_2_1}
\end{eqnarray}
}

Next, we show that $D(p\|q) \leq |\mathcal{X}|\log(k+1)$. This bound is a consequence of the method of types and its relation to exchangeability. For vector $\mathbf{x} \in \mathcal{X}^k$, let its type class be denoted as
\begin{equation}
\mathcal{T}_\mathbf{x}^{(k)} = \{ \mathbf{s} \in \mathcal{X}^k \ | \ \hat{p}_\mathbf{s} = \hat{p}_\mathbf{x} \},
\end{equation}
where the notation $\hat{p}_\mathbf{s}$ is taken from \eqref{eq:empirical_urn}. 

The idea is to lower bound $q(\mathbf{x})$
by restricting attention to the sequences within $\mathcal{T}_\mathbf{x}^{(k)}$: 
\begin{align}
    q(\mathbf{x}) &= \sum_{\mathbf{s}\in\mathcal{X}^k} p(\mathbf{s}) \left( \prod_{i=1}^{k} \hat{p}_\mathbf{s}(x_i) \right) \label{eq:q_x_types_1} \\
    &\geq \sum_{\mathbf{s} \in \mathcal{T}_\mathbf{x}^{(k)}}p(\mathbf{s}) \left( \prod_{i=1}^{k} \hat{p}_\mathbf{s}(x_i) \right) \label{eq:q_x_types_2} \\
    &= \sum_{\mathbf{s} \in \mathcal{T}_\mathbf{x}^{(k)}} p(\mathbf{x})  \left( \prod_{i=1}^{k} \hat{p}_\mathbf{x}(x_i) \right) \label{eq:q_x_types_3} \\
    &= |\mathcal{T}_{\mathbf{x}}| \; p(\mathbf{x})  \prod_{i=1}^{k} \hat{p}_\mathbf{x}(x_i), \label{eq:q_x_types}
\end{align}
where 
\begin{itemize}
\item \eqref{eq:q_x_types_1} is due to the urn codebook construction, with each term in the summation being the probability that a codeword $\mathbf{x}$ is generated from some $\mathbf{s} \in \mathcal{X}^k$ using its empirical distribution; 
\item \eqref{eq:q_x_types_3} follows because $p(\cdot)$ is exchangeable, so
due to the definition of the type class, 
$p(\mathbf{x}) = p(\mathbf{s}), \forall \mathbf{s} \in \mathcal{T}_\mathbf{x}^{(k)}$. 
\end{itemize}

The product distribution in~\eqref{eq:q_x_types} can be shown to be equal to $2^{- k H(\hat{p}_\mathbf{x})}$ as follows
\begin{align}
    \prod_{i=1}^{k} \hat{p}_\mathbf{x}(x_i) &= \prod_{c \in \mathcal{X}} \left(\hat{p}_\mathbf{x}(c)\right)^{N_\mathbf{x}(c) } \\ & = \prod_{c \in \mathcal{X}} \left(\hat{p}_\mathbf{x}(c)\right)^{k \hat{p}_\mathbf{x}(c) } \\
    &= 2^{k \sum_{c \in \mathcal{X}} \hat{p}_\mathbf{x}(c) \log \hat{p}_\mathbf{x}(c)} \\ & = 2^{- k H(\hat{p}_\mathbf{x})},
\label{eq:product_types}
\end{align}
where $N_\mathbf{x}(c)$ is the number of occurrences of $c$ in $\mathbf{x}$. 

Furthermore, by \cite[Theorem 11.1.3]{thegoat}, 
the total number of sequences in $\mathcal{T}_{\mathbf{x}}$ must satisfy 
\begin{equation}
|\mathcal{T}_{\mathbf{x}}| \geq \frac{1}{(k+1)^{|\mathcal{X}|}}2^{kH(\hat{p}_\mathbf{x})}.
\label{eq:method_of_types}
\end{equation} 
Combining \eqref{eq:q_x_types}, \eqref{eq:product_types} and \eqref{eq:method_of_types}, it follows that
\begin{equation}
    q(\mathbf{x}) \geq \frac{1}{(k+1)^{|\mathcal{X}|}} p(\mathbf{x}).
\end{equation}
This implies that 
\begin{equation}
D(p\|q) \leq |\mathcal{X}|\log(k+1).
\label{eq:theorem_2_2}
\end{equation}
Together, \eqref{eq:theorem_2_1_2}, \eqref{eq:theorem_2_1} and \eqref{eq:theorem_2_2} give \eqref{eq:theorem_2}.

\section{Proof of Theorem \ref{thm:exch_extend}}
\label{app:exch_extend}


To simplify notation, let $\tilde{p}(x_1, \cdots, x_d)$ denote the distribution of the extended sequence $(X_1, \cdots, X_d)$. For $\mathbf{s} \in \mathcal{X}^d$, we let $h_\mathbf{s}(x_1, \cdots, x_k)$ denote the distribution of $k$ draws without replacement from an urn containing the elements of $\mathbf{s}$ and let $m_\mathbf{s}(x_1, \cdots, x_k)$ denote the distribution of $k$ draws with replacement from the same urn. Note that $m_\mathbf{s}(x_1, \cdots, x_k)$ is equivalent to the i.i.d.\ distribution with the empirical distribution of $\mathbf{s}$ as its marginal.

Using the extended distribution $\tilde{p}$, the distribution of the sources is equivalent to the following weighted sum of urn distributions:
\begin{equation}
\label{eq:weighted-h}
    p(\mathbf{x}) = \sum_{\mathbf{s} \in \mathcal{X}^d} \tilde{p}(\mathbf{s}) h_\mathbf{s}(\mathbf{x}).
\end{equation}
This follows from exchangeability, since the distribution of any $k$ distinct elements of $(X_1, \cdots, X_d)$ has the same distribution as $(X_1, \cdots, X_k)$. 

If instead we sample with replacement, then we would have the distribution
\begin{equation}
\label{eq:weighted-m}
    q(\mathbf{x}) = \sum_{\mathbf{s} \in \mathcal{X}^d} \tilde{p}(\mathbf{s}) m_\mathbf{s}(\mathbf{x}).
\end{equation}
This corresponds to the distribution of the extended urn codebook. The goal is to show that $D(p\|q)$ is small.



Using the log-sum inequality, the KL divergence can be upper bounded as
\begin{align}
\label{eq:log-sum}
D(&p\|q) 
 =  \sum_{\mathbf{x} \in \mathcal{X}^k} p(\mathbf{x}) \log \left( \frac{p(\mathbf{x})}{q(\mathbf{x})} \right) \\
&=   \sum_{\mathbf{x} \in \mathcal{X}^k} \left( \sum_{\mathbf{s} \in \mathcal{X}^d} \tilde{p}(\mathbf{s})h_\mathbf{s}(\mathbf{x}) \right) \log \left( \frac{\sum_{\mathbf{s} \in \mathcal{X}^d} \tilde{p}(\mathbf{s})h_\mathbf{s}(\mathbf{x})}{\sum_{\mathbf{s} \in \mathcal{X}^d} \tilde{p}(\mathbf{s})m_\mathbf{s}(\mathbf{x})} \right) \\
&\leq  \sum_{\mathbf{s} \in \mathcal{X}^d} \tilde{p}(\mathbf{s}) \sum_{\mathbf{x} \in \mathcal{X}^k} h_\mathbf{s}(\mathbf{x}) \log \left( \frac{h_\mathbf{s}(\mathbf{x})}{m_\mathbf{s}(\mathbf{x})} \right) \\
&=  \sum_{\mathbf{s} \in \mathcal{X}^d} \tilde{p}(\mathbf{s}) D(h_\mathbf{s}\|m_\mathbf{s}).
\end{align}
From \cite[Theorem 4.5]{matus}, we know that for all $\mathbf{s} \in \mathcal{X}^d$,
\begin{equation}
\begin{split}
        D(h_\mathbf{s}\|m_\mathbf{s}) \leq (|\mathcal{X}|-1) \log \left(\frac{d-1}{d-k} \right).
\end{split}
\end{equation}
Therefore, $D(p\|q)$ is bounded from above by the same quantity. {\color{black}This line of reasoning is similar to the reasoning used in \cite{exch-it-4} to prove \eqref{eq:stam-arg}. }

{\color{black}
Next, we apply a similar line of reasoning as the proof of Theorem \ref{thm:exch} by considering collision-free sampling patterns. We represent the sampling process by defining the positions of $\mathbf{s}$ that are sampled as $\mathbf{w} = (w_1, \cdots, w_k)$ where each $w_i$ is i.i.d.\ over $[d]$.
Likewise, we define the set of all collision-free sample patterns as
$\mathcal{F} = \{ \mathbf{w} \in [d]^k \ | \ w_i \neq w_j \text{ for all } i \neq j \}$. It follows that
\begin{align}
    q(\mathbf{x}) &= 
\sum_{\mathbf{s} \in \mathcal{X}^k} 
\sum_{\mathbf{w} \in [d]^k} 
p(\mathbf{s}) 
p(\mathbf{w}) 
\mathbbm{1}({\mathbf{x}=\mathbf{s}_\mathbf{w}}) \\ 
    &\geq 
\sum_{\mathbf{s} \in \mathcal{X}^k} 
\sum_{\mathbf{w} \in \mathcal{F}}
p(\mathbf{s}) 
p(\mathbf{w}) 
\mathbbm{1}({\mathbf{x}=\mathbf{s}_\mathbf{w}}) \\ 
    &= \sum_{\mathbf{w} \in \mathcal{F}} \frac{1}{d^k}p(\mathbf{x}) \\
    & = \frac{d^{\underline{k}}}{d^k}p(\mathbf{x}).
\end{align}
Therefore, we have that
\begin{equation}
    D(p\|q) \leq \log \left( \frac{d^k}{d^{\underline{k}}} \right).
\end{equation}
}

\section{Converse for Scheduling}
\label{app:converse_scheduling}

We prove \eqref{eq:converse_scheduling} using a volume bound argument on the total number of possible scheduling patterns in relation to the maximum number of scheduling patterns can correspond to a single codeword. This allows us to show that the optimal common message length is bounded from below as \eqref{eq:converse_scheduling}, i.e.,
\begin{equation}
    R^* \geq k\log(b) - \log \left( \frac{n^k}{n^{\underline{k}}} \right).
\end{equation}


We begin by considering the total number of scheduling patterns. This amounts to the number of ways to select $k$ users out of $n$ users, then assigning them each a distinct value from $[b]$. By a simple counting argument, there are a total of
\begin{equation}
    {{n}\choose{k}}{{b}\choose{k}}k!
\end{equation}
possible scheduling patterns.

Next, we characterize $d_{\text{max}}$ defined as the maximum number of scheduling patterns that a single codeword can correspond to. Let $\mathbf{v} = \left[v_1, \ldots, v_b \right]^{\sf T}$, where $v_{\ell}$ is the number of times $\ell$ occurs in the entries of the codeword. We have $\sum_{\ell=1}^{b} v_{\ell} = n$. The total number of different scheduling patterns that can correspond to this codeword is
\begin{equation}
    \sum_{\mathcal{U} \in {{[b]}\choose{k}}} \prod_{\ell \in \mathcal{U}} v_\ell
\end{equation} 
where ${{[b]}\choose{k}}$ is the set of all size-$k$ subsets of $[b]$, and the product corresponds to the number of ways to pick one user from each set of $v_\ell$ entries where $\ell$ occurs in the codeword.
To find $d_{\text{max}}$, we optimize over $\mathbf{v}$
\begin{align}
    \mathbf{v}^* & = \argmax_{\mathbf{v}} \sum_{\mathcal{U} \in {{[b]}\choose{k}}} \prod_{\ell \in \mathcal{U}} v_\ell \\ &= \argmax_{\mathbf{v}} \sum_{\mathcal{U} \in {{[b]}\choose{k}}} k!\prod_{\ell \in \mathcal{U}} \frac{v_\ell}{n}.
\end{align}
The objective on the right-hand side can be interpreted as the probability that $k$ i.i.d.\ samples from the distribution $\left( \frac{v_1}{n}, \cdots, \frac{v_b}{n} \right)$ have distinct values. In \cite{dia-fre}, it is shown that this probability is maximized by the uniform distribution over $[b]$. This shows that the optimal $v_\ell = \frac{n}{b}$ for all $\ell \in [b]$ and
\begin{equation}
    d_{\text{max}} = {{b}\choose{k}} \left(\frac{n}{b}\right)^k.
\end{equation}

Let $T$ denote the output of an encoder $f$ designed to communicate a particular schedule without error to the $k$ active users out of a total of $n$ users. Since there are ${{n}\choose{k}}{{b}\choose{k}}k!$ ways to schedule $k$ users into $b$ slots and that a single codeword can only cover $d_{\text{max}}$ different schedules, we have that
\begin{equation}
\label{p_t_scheduling}
    \mathrm{Pr}(T = t) \leq \frac{d_\text{max}}{{{n}\choose{k}}{{b}\choose{k}}k!} = \frac{ n^{\underline{k}}}{b^k n^k}.
\end{equation}
Therefore
\begin{align}
    H(T) &= - \sum_{t} \mathrm{Pr}(T = t) \log(\mathrm{Pr}(T = t)) \\
    &\geq - \sum_{t} \mathrm{Pr}(T = t) \log  \left( \frac{n^{\underline{k}}}{b^k n^k} \right) \\ 
    &= k\log(b) - \log \left( \frac{n^k}{n^{\underline{k}}} \right).
\end{align}

\section{Converse for Categorization}
\label{app:converse_categorization}

We use a volume bound argument to show \eqref{eq:converse_categorization}, i.e. for the problem of categorization, the optimal common message length is bounded from below as
\begin{equation}
    R^* \geq kH(\rho) - \log \left( \frac{n^k}{n^{\underline{k}}} \right),
\end{equation}
where $\rho = \left( \frac{k_1}{k}, \cdots, \frac{k_c}{k} \right)$.

We begin by characterizing the number of ways $k$ out of $n$ users can be categorized. This amounts to choosing $k$ users amongst $n$, then splitting and assigning them into $c$ categories, where the sizes of the categories are fixed to be $k_\ell$ for all $\ell \in [c]$. By simple counting, there are a total of
\begin{equation}
    {{n}\choose{k}} {{k}\choose{k_1 \cdots k_c}}
\end{equation}
categorization patterns.

Next, we upper bound $d_{\text{max}}$, which is defined to be the maximum number categorization patterns that a single codeword can correspond to. 
Let $\mathbf{v} = \left[v_1, \ldots, v_b \right]^{\sf T}$, where $v_{\ell}$ is the number of times $\ell$ occurs in the entries of the codeword. We have $\sum_{\ell=1}^{b} v_{\ell} = n$. 
Then, the number of different categorization patterns covered by this codeword is
\begin{equation}
    \prod_{\ell=1}^{c} {{v_{\ell}}\choose{k_{\ell}}} \leq \prod_{\ell=1}^{c} \frac{v_{\ell}^{k_{\ell}}}{k_{\ell}!}.
\end{equation}
To upper bound $d_\text{max}$, we optimize over $\mathbf{v}$. Notice that 
\begin{equation}
    \argmax_{\mathbf{v}} \prod_{\ell=1}^{c} \frac{v_{\ell}^{k_{\ell}}}{k_{\ell}!} = \argmin_{\mathbf{v}} -\sum_{\ell=1}^{c} \frac{k_\ell}{k} \log\left(\frac{v_{\ell}}{n}\right).
\end{equation}
The objective of the right hand side is equivalent to minimizing the KL divergence between $\left( \frac{k_1}{k}, \cdots, \frac{k_c}{k} \right)$ and $\left( \frac{v_1}{n}, \cdots, \frac{v_c}{n} \right)$. Therefore, we should set $v^{*}_{\ell} = \frac{nk_{\ell}}{k}$ which yields the upper bound
\begin{equation}
\label{eq:ub_dmax}
d_{\text{max}} \leq \prod_{\ell=1}^{c} \frac{\left( \frac{nk_{\ell}}{k} \right)^{k_{\ell}}}{k_{\ell}!}.
\end{equation}

Let $T$ denote the output of an encoder $f$ designed to communicate a particular categorization without error to the $k$ active users out of a total of $n$ users. Since there are ${{n}\choose{k}} {{k}\choose{k_1 \cdots k_c}}$ ways to categorize $k$ users and that a single codeword can only cover $d_{\text{max}}$ different categorizations, we have that
\begin{equation}
\label{p_t_categorization}
    \mathrm{Pr}(T = t) \leq \frac{d_\text{max}} {{{n}\choose{k}} {{k}\choose{k_{1} \cdots k_{c}}}}.
\end{equation}
Therefore,
\begin{align}
    H(T) &= - \sum_{t} \mathrm{Pr}(T = t) \log(\mathrm{Pr}(T = t)) \\
    &\geq - \sum_{t} \mathrm{Pr}(T = t) \log  \left( \frac{d_\text{max}} {{{n}\choose{k}} {{k}\choose{k_{1} \cdots k_{c}}}} \right) \\ 
    &=\log \left( {{n}\choose{k}} {{k}\choose{k_{1} \cdots k_{c}}}\frac{1}{d_\text{max}}\right) \\
    &\geq kH(\rho) - \log \left( \frac{n^k}{n^{\underline{k}}} \right).
\end{align}

\section{Proof of Theorem \ref{thm:random_users}}
\label{app:random_users}

The proof is similar to that of Theorem \ref{thm:main}. We use the same codebook as in Section \ref{section:lossless} and also the same decoders. For the encoder, we modify it slightly to accommodate a varying number of users. 

For a codebook $\mathbf{m} = (\mathbf{c}^{(1)}, \mathbf{c}^{(2)}, \cdots )$, define
\begin{eqnarray}
\label{eq:encoder_random_K}
g_{\mathbf{m}}(x_1,\cdots x_K, a_1,\cdots a_K) = & \min &  t \\
& \mathrm{s.t.} & {c^{(t)}_{a_i} = x_i , \ \forall i \in [K]}. \nonumber
\end{eqnarray}
    Let $T = g_\mathbf{M}(X_1, \cdots, X_K, A_1, \cdots, A_K)$ be the index of the first matching codeword 
    using randomly generated codebook $\mathbf M$. Since the sources are distributed i.i.d.\ according to $p(x)$, we choose to generate the entries of the codewords in $\mathbf{M}$ in an i.i.d.\ fashion according to $p(x)$ as well. By the same argument as the proof of Theorem \ref{thm:main}, we have that
    \begin{equation}
        \mathbb{E}[\log(T)|K=k] \leq kH(p).
    \end{equation}
    Therefore,
    \begin{equation}
        \mathbb{E}[\log(T)] \leq \mathbb{E}[K]H(p).
    \end{equation}
    Combining this with \eqref{eq:max-entropy-arg}, we have that
    \begin{equation}
        H(T) \leq \mathbb{E}[K]H(p) + \log(\mathbb{E}[K]H(p) + 1) + 1.
    \end{equation}
    Therefore,
    \begin{equation}
        R^* < \mathbb{E}[K]H(p) + \log(\mathbb{E}[K]H(p) + 1) + 2.
    \end{equation}

\section{Proof of Theorem \ref{thm:definetti_upper}}

\label{app:definitti_upper}
    In order to lower bound the KL divergence $D(r\| q)$, we first lower bound the cross-entropy using Jensen's inequality, i.e.,
    \begin{align}
    \label{eq:cross-entropy}
	    \sum_{\mathbf{x} \in [d]^k} r(\mathbf{x}) \log \left( \frac{1}{q(\mathbf{x})} \right) &= \mathbb{E} \left[ \log \left( \frac{1}{q(\mathbf{x})} \right) \right] \\
        &\geq \log \left( \frac{1}{\mathbb{E}[q(\mathbf{x})]} \right).
    \end{align}
    It remains to establish an upper bound on $\mathbb{E}[q(\mathbf{x})]$. To do so, we take inspiration from \cite{dia-fre}. Let 
    \begin{equation}
        \mathcal{B} = \{ \mathbf{x} \in [d]^k \ | \ x_i = x_j \ \text{for some } i \neq j \}.
    \end{equation}
    By definition of $r$, we know that $r(\mathbf{x}) = 0$ for all $\mathbf{x} \in \mathcal{B}$. Therefore
    \begin{equation}
        \mathbb{E}[q(\mathbf{x})] = \sum_{\mathbf{x} \in \mathcal{B}^{\sf c}} r(\mathbf{x})q(\mathbf{x}) = \left({{d}\choose{k}}k!\right)^{-1} \sum_{\mathbf{x} \in \mathcal{B}^{\sf c}} q(\mathbf{x}).
    \end{equation}
    From \cite{dia-fre}, we know that the i.i.d.\ mixture distribution that maximizes $\sum_{\mathbf{x} \in \mathcal{B}^{\sf c}} q(\mathbf{x})$ is the uniform distribution over $[d]^k$. Therefore
    \begin{equation}
        \mathbb{E}[q(\mathbf{x})] \leq \left({{d}\choose{k}}k!\right)^{-1} 
        \sum_{\mathbf{x} \in \mathcal{B}^{\sf c}} \frac{1}{d^k} = \left({{d}\choose{k}}k!\right)^{-1}  \frac{d^{\underline{k}}}{d^k}.
    \end{equation}
    Combining the above bounds, we have that
    \begin{align}
       D(r\|q) &= \mathbb{E} \left[ \log \left( \frac{1}{q(\mathbf{x})} \right) \right] - H(\mathbf{X}) \\
       &\geq \log \left({{d}\choose{k}}k!\right) + \log \left( \frac{d^{k}}{d^{\underline{k}}} \right) - H(\mathbf{X}) \\
       &= \log \left( \frac{d^{k}}{d^{\underline{k}}} \right),
    \end{align}
    where $\mathbf{X} \sim r(\mathbf{x})$.

\bibliographystyle{IEEEtran}
\bibliography{references}

\end{document}